\documentclass[floatfix,twocolumn,showpacs,prl,amsmath]{revtex4}
\usepackage{graphicx}
\usepackage{bm}
\usepackage{amssymb}
\usepackage{dcolumn}
\usepackage{subfigure}
\usepackage{threeparttable}
\usepackage{multirow}
\usepackage[usenames,dvipsnames]{color}
\usepackage{mathtools}

\begin{document}
\newcommand{\vn}[1]{{\boldsymbol{#1}}}
\newcommand{\vht}[1]{{\boldsymbol{#1}}}
\newcommand{\matn}[1]{{\bf{#1}}}
\newcommand{\matnht}[1]{{\boldsymbol{#1}}}
\newcommand{\bege}{\begin{equation}}
\newcommand{\ee}{\end{equation}}
\newcommand{\bal}{\begin{aligned}}
\newcommand{\defbar}{\overline}
\newcommand{\SM}{\scriptstyle}
\newcommand{\eal}{\end{aligned}}
\newcommand{\torkance}{t}
\newcommand{\udot}{\overset{.}{u}}
\newcommand{\exponential}[1]{{\exp(#1)}}
\newcommand{\phandot}[1]{\overset{\phantom{.}}{#1}}
\newcommand{\phandag}{\phantom{\dagger}}
\newcommand{\Trace}{\text{Tr}}
\newcommand{\Bxc}{\Omega}
\setcounter{secnumdepth}{2} 
\title{Direct and inverse spin-orbit torques}
\author{Frank Freimuth}
\email[Corresp.~author:~]{f.freimuth@fz-juelich.de}
\author{Stefan Bl\"ugel}
\author{Yuriy Mokrousov}
\affiliation{Peter Gr\"unberg Institut and Institute for Advanced Simulation,
Forschungszentrum J\"ulich and JARA, 52425 J\"ulich, Germany}
\date{\today}
\begin{abstract}
In collinear magnets lacking inversion symmetry application of
electric currents induces torques on the magnetization and
conversely magnetization dynamics induces electric currents.
The two effects, which both rely on spin-orbit interaction (SOI), 
are reciprocal to each other and denoted direct spin-orbit torque
(SOT)
and inverse spin-orbit torque (ISOT), respectively. We derive
expressions for SOT and ISOT within the Kubo linear response
formalism.
We show that expressions suitable for density-functional theory
calculations can be derived either starting from a Kohn-Sham
Hamiltonian
with time-dependent exchange field or by expressing general
susceptibilities in terms of the Kohn-Sham susceptibilities.
For the case of magnetic bilayer systems we derive the
general form of the ISOT current induced under ferromagnetic resonance.
Using \textit{ab initio} calculations within density-functional theory 
we investigate
SOT and ISOT in Co/Pt(111) magnetic bilayers. 
We determine the spatial distribution of spin and charge
currents as well as torques in order to expose the
mechanisms underlying SOT and ISOT and to highlight their
reciprocity on the microscopic level. We find that the spin Hall
effect
is position-dependent close to interfaces.  
\end{abstract}

\pacs{72.25.Ba, 72.25.Mk, 71.70.Ej, 75.70.Tj}

\maketitle
\section{Introduction}

In ferromagnetic materials Faraday's law of induction needs to be
generalized to include so-called spinmotive forces, 
i.e., electric fields
induced by the magnetization 
dynamics~\cite{josephson_effect_ferromagnets,
linear_momentum_ferromagnetics_volovik,
generalized_faraday_barnes_maekawa}.
The spinmotive force can be interpreted as the 
reciprocal of the current-induced torque: A moving domain wall induces
a spinmotive force and conversely an applied current drives domain wall
motion. 
Thus, the electric fields induced by magnetization dynamics 
generate a feedback effect on the magnetization 
via the current-induced
torques which they 
produce~\cite{self_consistent_spin_transport_magnetization_dynamics}.

Spinmotive forces do not only occur in noncollinear magnetic structures
such as 
domain-walls~\cite{universal_electromotive_force_induced_by_dwm}  
and skyrmions~\cite{emergent_electrodynamics_skyrmions}
but can arise also in collinear magnets due to the interplay 
of spin orbit interaction (SOI) with bulk or structural 
inversion asymmetry~\cite{prediction_giant_spin_motive_rashba,
spin_motive_rashba_strong_sd_coupling_regime}.
Spin-orbit torques (SOTs)~\cite{torque_macdonald,
manchon_zhang_2009,quantum_kinetic_rashba_macdonald,
current_induced_torques_rashba_ferromagnets,
quantum_kinetic_rashba_manchon,
semi_classical_modelling_stiles,
current_induced_torques_haney,
phenomenology_hals_brataas,
magnetization_dynamics_rashba},
i.e., current-induced torques 
originating from SOI
in inversion asymmetric collinear magnets,
are the reciprocal to the electric fields induced 
by magnetization dynamics in collinear 
magnets~\cite{charge_pumping_Ciccarelli,spin_motive_force_hals_brataas}. 
Thus, we will denote the latter as inverse spin-orbit 
torques (ISOTs) in
the following. ISOTs constitute a special case of
spinmotive forces. 

While earlier experiments on SOTs estimated the current-induced torques
indirectly from the onset of nucleation 
of reversed domains~\cite{CoPtAlO_spin_torque_rashba_Gambardella} 
or magnetization switching at critical
current densities~\cite{CoPtAlO_perpendicular_switching_Gambardella,
current_induced_switching_using_spin_torque_from_spin_hall_buhrman,
spin_torque_switching_TaCoFeB_Buhrman}
direct measurements of SOTs have been performed
recently in bilayer systems 
and the SOT has been determined
as a function of magnetization
direction $\hat{\vn{M}}$~\cite{symmetry_spin_orbit_torques,
layer_thickness_dependence_current_induced_effective_field_TaCoFeB_Hayashi,
angular_temperature_dependence_sot_ta}.
Two qualitatively different SOT components are found in these
experiments on bilayer systems, the first one is an
even function of $\hat{\vn{M}}$, the second 
one is an odd function.
Denoting the applied in-plane electric field by $\vn{E}$
and the unit vector in the out-of-plane 
direction by $\hat{\vn{e}}_{z}$, they are given by
$\vn{T}^{\rm even}={T}^{\rm even}
\hat{\vn{M}}\times[(\hat{\vn{e}}_{z}\times\vn{E})\times\hat{\vn{M}}]$
and $\vn{T}^{\rm odd}=
{T}^{\rm odd}(\hat{\vn{e}}_{z}\times\vn{E})\times\hat{\vn{M}}$
to lowest order in $\hat{\vn{M}}$.

In bilayer systems based on 5d transition metals
with large spin Hall effect (SHE), such as
AlO$_{x}$/Co/Pt,
MgO/CoFeB/Ta
and CoFeB/W,
the dominant contribution to $\vn{T}^{\rm even}$
arises from the 
SHE~\cite{current_induced_switching_using_spin_torque_from_spin_hall_buhrman,
spin_torque_switching_TaCoFeB_Buhrman,
stt_devices_giant_she_tungsten,
domain_wall_depinning_haazen,
chiral_domain_wall_motion_parkin,
chiral_domain_wall_motion_beach,
ibcsoit}. 
Conversely, in Ni$_{80}$Fe$_{20}$/Pt the spin current
pumped into Pt by exciting the ferromagnetic resonance (FMR)
of Ni$_{80}$Fe$_{20}$ induces an electric field via the inverse spin Hall
effect (ISHE)~\cite{gilbert_damping_ultrathin_films,
enhanced_gilbert_damping_ferromagnetic_films,
prl_mosendz_spin_pumping}.
Rashba SOI provides an important contribution to $\vn{T}^{\rm odd}$ in
these
bilayer systems~\cite{current_induced_torques_haney,semi_classical_modelling_stiles}.
Due to the reciprocity between SOT and ISOT, an 
additional ISOT is expected as well from the
Rashba SOI at the bilayer 
interface~\cite{prediction_giant_spin_motive_rashba,
spin_motive_rashba_strong_sd_coupling_regime}.
This theoretical prediction, that the ISOT 
in bilayer systems should not arise purely from the combination
of spin pumping and ISHE, is supported by the experimental
observation that for the reciprocal phenomenon, the SOT,
$\vn{T}^{\rm odd}$ can be as large as or 
even larger than $\vn{T}^{\rm even}$~\cite{symmetry_spin_orbit_torques,
layer_thickness_dependence_current_induced_effective_field_TaCoFeB_Hayashi,
angular_temperature_dependence_sot_ta}.

So far only the dc voltage due to 
FMR-driven ISOT has been studied
intensively in bilayer systems~\cite{prl_mosendz_spin_pumping,
prb_mosendz_spin_pumping,prl_czeschka_spin_pumping,
spin_pumping_amr_bilayers,
experimental_test_spin_mixing}.
However, after the 
theoretical 
prediction~\cite{ac_voltage_spin_pumping_bauer}
that the ac component is expected
to be much larger than the dc one,
several recent experiments have been 
devoted to its 
measurement~\cite{ac_spin_hall,
phase_sensitive_detection_spin_pumping_ac_inverse_she,
detection_microwave_spin_pumping_ishe}.
As will be discussed in this work
it is expected 
from the reciprocity of ISOT and SOT 
that
the dc voltage generated by the FMR-driven ISOT is 
proportional to $\vn{T}^{\rm even}$,
while the ac voltage is determined 
by both $\vn{T}^{\rm even}$ and $\vn{T}^{\rm odd}$.
Since the ac voltages associated with $\vn{T}^{\rm even}$ 
and $\vn{T}^{\rm odd}$ exhibit a phase 
difference of $\pm 90^{\circ}$ a
non-trivial phase relationship between ac 
signal and magnetization trajectory is expected.
Phase-sensitive measurements of the ac ISOT-signal induced
under FMR can thus be complementary to experiments 
on the SOT phenomenon. Both types of experiments, i.e.,
measuring the induced voltage under FMR on the one hand
and measuring on the other hand the current-induced
torque on the magnetization,
can thus serve to determine 
$\vn{T}^{\rm even}$ and $\vn{T}^{\rm odd}$ and from
them
the parameters needed to model them, 
notably spin-diffusion length,
spin-mixing conductance, SHE-angle as well as
Rashba and Dresselhaus parameters.  

This article is organized as follows: 
In Sec.~\ref{sec_relate_direct_and_inverse_sot} we discuss the 
Kubo formalism expressions for both SOT and ISOT. 
In the case of the SOT phenomenon,
the torque on the magnetization is given by $\vn{T}=\vn{\torkance}\vn{E}$,
which defines the torkance tensor $\vn{\torkance}$.
We show that also the ISOT can be captured conveniently in terms 
of $\vn{\torkance}$, which is a consequence of the 
reciprocity between
SOT and ISOT. 
In Sec.~\ref{sec_completing_response_matrix} we show that expressions for both 
ISOT and Gilbert damping can be derived consistently
based on Kohn-Sham theory with a time-dependent exchange field.
In Sec.~\ref{sec_many_electron_resp} we show that these
expressions
can also be obtained by expressing general many-body susceptibilities
in terms of the corresponding Kohn-Sham susceptibilities.
Exploiting the reciprocity between SOT and ISOT we then predict 
in 
Sec.~\ref{sec_angular_dependence_sot_isot}
the angular
dependence of ISOT in magnetic bilayers 
from the angular dependence of SOT
recently measured in these systems. 
In particular we derive and discuss
the FMR-induced currents for 
various magnetization directions in 
bilayer systems. 
In Sec.~\ref{sect_consequences_reciprocity} a minimal model to
describe even SOT and ISOT in bilayers is discussed.
In Sec.~\ref{sect_odd_boltzmann} we consider odd SOT and ISOT
within the Boltzmann formalism.
In Sec.~\ref{sec_first_principles} we investigate SOT and ISOT
for a magnetic bilayer composed of a Co layer on Pt(111). 
Computing spin currents, ISOT-induced charge currents 
and torkances layer-resolved we make contact with 
phenomenological models and extract model parameters.
We conclude by a summary in Sec.~\ref{sec_summary}.
\section{Relationship between direct SOT and inverse SOT}
\label{sec_relate_direct_and_inverse_sot}
\subsection{Induced currents under time-dependent magnetization}
Reciprocity between current-induced torques and spinmotive forces
has been discussed in detail in the framework of 
phenomenological 
modelling~\cite{effects_nonadiabaticity_voltage_moving_domain_wall,
she_phenomenology_magnetic_dynamics,spin_motive_force_hals_brataas,
charge_pumping_Ciccarelli}.
In this section, we revisit this reciprocity on the basis of the Kubo linear
response formalism, which is well-suited to study SOT and ISOT from
first principles. 

Within the local spin density approximation (LSDA) the interacting
many-electron system is described by an effective 
single-particle Hamiltonian of the form 
\bege\label{eq_time_dependent_hamil}
H(\vn{r},t)=H_{0}(\vn{r})-\vn{m}\cdot\hat{\vn{M}}(t)\Bxc^{\rm xc}(\vn{r}),
\ee
where the time-independent $H_{0}$ contains kinetic energy, 
scalar potential and SOI, while the second term on the right-hand side
describes the exchange interaction. $\hat{\vn{M}}(t)$ is a normalized
vector which points in the direction of magnetization. In order to
describe the electronic system at the ferromagnetic resonance we
assume that $\hat{\vn{M}}(t)$ is precessing.
The time-dependence of the Hamiltonian arises
from this precession of magnetization. $\vn{m}=-\mu_{\rm B}\vht{\sigma}$ with the
Bohr magneton $\mu_{\rm B}$  and the vector of Pauli 
spin matrices  $\vht{\sigma}=(\sigma_x,\sigma_y,\sigma_z)^{\rm T}$ is 
the spin magnetic moment operator. 
$\Bxc^{\rm xc}(\vn{r})$ is the exchange field, i.e., the
difference between the potentials of majority and 
minority electrons $\Bxc^{\rm xc}(\vn{r})=\frac{1}{2\mu_{\rm B}}\left(V^{\rm eff}_{\rm minority}(\vn{r})-V^{\rm eff}_{\rm majority}(\vn{r}) \right)$.
Around the time $t$ we can approximate the motion of $\hat{\vn{M}}$ by
\bege
\hat{\vn{M}}(t\!+\!\Delta t)\!-\!\hat{\vn{M}}(t)
\simeq
\frac{d\hat{\vn{M}}(t)}{dt}\Delta t
\simeq
\frac{d\hat{\vn{M}}(t)}{dt}\frac{\sin(\omega \Delta t)}{\omega}
\ee
for small time changes $\Delta t$ and a small but arbitrary 
frequency $\omega$ with $\omega \Delta t\ll 1$.
Likewise, the Hamiltonian can be approximated as
\bege
H(\vn{r},t+\Delta t)\simeq H(\vn{r},t)
-
\vn{m}\cdot\frac{d\hat{\vn{M}}(t)}{dt}
\Bxc^{\rm xc}(\vn{r})
\frac{\sin(\omega \Delta t)}{\omega}.
\ee 
The $\Delta t$-dependent term 
\begin{gather}
\begin{aligned}
\label{eq_time_dependent_perturbation}
&V(\vn{r},\Delta t)=
-
\vn{m}\cdot\frac{d\hat{\vn{M}}(t)}{dt}
\Bxc^{\rm xc}(\vn{r})
\frac{\sin(\omega \Delta t)}{\omega}\\
&=
-\vn{m}\!\!\cdot\!\!
\left[
\hat{\vn{M}}(t)
\!\times\!
\left(\!
\frac{d\hat{\vn{M}}(t)}{dt}
\!\times\!
\hat{\vn{M}}(t)
\!\right)
\right]\Bxc^{\rm xc}(\vn{r})
\frac{\sin(\omega \Delta t)}{\omega}\\
&=
\frac{\sin(\omega \Delta t)}{\omega}
\left(
\hat{\vn{M}}(t)
\times
\frac{d\hat{\vn{M}}(t)}{dt}
\right)
\cdot
\vht{\mathcal{T}}(\vn{r},t)
\end{aligned}\raisetag{2\baselineskip}
\end{gather}
acts as a time-dependent perturbation on the 
eigenstates of $H(\vn{r},t)$. 
Here, $\vht{\mathcal{T}}(\vn{r},t)=\vn{m}\times \hat{\vn{M}}(t)\Bxc^{\rm xc}(\vn{r})$ is the torque operator.

Within linear response the current density 
in $\alpha$ direction, $j_{\alpha}$, induced by the 
time-dependent perturbation 
Eq.~\eqref{eq_time_dependent_perturbation} is 
given by
\bege\label{eq_induced_current}
j_{\alpha}(t)\!=\!\!
\sum_{\beta}\!
\frac{e}{V}\!
\lim_{\omega\to 0}\!
\frac{{\rm Im}G_{v_{\alpha}^{\phantom{\alpha}},\mathcal{T}_{\beta}^{\phantom{\alpha}}}^{\rm R}
(\hbar\omega,\hat{\vn{M}}(t))}{\hbar\omega} 
\!\left(\!\!
\hat{\vn{M}}(t)
\!\!\times\!\!
\frac{d\hat{\vn{M}}(t)}{dt}
\!\right)_{\beta},
\ee
where $e>0$ is the elementary positive charge, $V$ is the volume and
$G_{v_{\alpha}^{\phantom{\alpha}},\mathcal{T}_{\beta}^{\phantom{\alpha}}}^{\rm R}(\hbar\omega,\hat{\vn{M}})$ is 
the Fourier transform of 
the retarded velocity-torque correlation function, i.e., 
\bege\label{eq_velocity_torque_correlation}
G_{v_{\alpha}^{\phantom{\alpha}},\mathcal{T}_{\beta}^{\phantom{\alpha}}}^{\rm R}(\hbar\omega,\hat{\vn{M}})=-i\int\limits_{0}^{\infty}dt e^{i\omega t}
\left\langle
[
v_{\alpha}(t),\mathcal{T}_{\beta}(0)
]_{-}
\right\rangle,
\ee
evaluated for the time-independent Hamiltonian
\bege\label{eq_static_Hamiltonian}
H_{\hat{\vn{M}}}(\vn{r})=
H_{0}(\vn{r})
-
\vn{m}
\cdot
\hat{\vn{M}}
\Bxc^{\rm xc}(\vn{r})
\ee
of a system with magnetization in 
direction $\hat{\vn{M}}=\hat{\vn{M}}(t)$.
Eq.~\eqref{eq_velocity_torque_correlation} describes the correlation
between the polar vector $\vn{v}$ and the axial vector
$\vn{\mathcal{T}}$.
This polar-axial correlation is nonzero only when inversion symmetry
is broken.

Next, we compare Eq.~\eqref{eq_induced_current} to 
the expressions describing SOTs.
Within linear response to an applied 
electric field $\vn{E}$ the SOT on the magnetization
is $\vn{T}(\hat{\vn{M}})=\vn{\torkance}(\hat{\vn{M}})\vn{E}$, 
where 
the torkance 
tensor $\vn{\torkance}(\hat{\vn{M}})$ is given by~\cite{ibcsoit,mothedmisot}
\bege\label{eq_torkance_tensor}
\torkance_{\alpha\beta}(\hat{\vn{M}})=-e 
\lim_{\omega\to 0} 
\frac{{\rm Im}G_{\mathcal{T}_{\alpha}^{\phantom{\alpha}}\!\!,v_{\beta}^{\phantom{\alpha}}}^{\rm R}(\hbar\omega,\hat{\vn{M}})}{\hbar\omega}
\ee
in terms of the Fourier transform of the retarded torque-velocity 
correlation function
\bege\label{eq_torque_velocity_correlation}
G_{\mathcal{T}_{\alpha}^{\phantom{\alpha}}\!\!,v_{\beta}^{\phantom{\alpha}}}^{\rm R}(\hbar\omega,\hat{\vn{M}})=
-i\int\limits_{0}^{\infty}dt e^{i\omega t}
\left\langle
[
\mathcal{T}_{\alpha}(t),v_{\beta}(0)
]_{-}
\right\rangle
\ee
of the system with Hamiltonian Eq.~\eqref{eq_static_Hamiltonian}.
 
The spectral densities of the Green functions defined in
Eq.~\eqref{eq_velocity_torque_correlation} and 
in Eq.~\eqref{eq_torque_velocity_correlation} are given by  
\bege
\begin{aligned}
S_{v_{\alpha}^{\phantom{\alpha}},\mathcal{T}_{\beta}^{\phantom{\alpha}}}(t,t',\hat{\vn{M}})
=&
\frac{1}{2\pi}
\left\langle
[
v_{\alpha}(t)
,
\mathcal{T}_{\beta}(t')
]_{-}
\right\rangle,\\
S_{\mathcal{T}_{\alpha}^{\phantom{\alpha}}\!\!,v_{\beta}^{\phantom{\alpha}}} (t,t',\hat{\vn{M}})
=&
\frac{1}{2\pi}
\left\langle
[
\mathcal{T}_{\alpha}(t)
,
v_{\beta}(t')
]_{-}
\right\rangle
\end{aligned}
\ee
and their Fourier transforms satisfy the relations
\bege
\begin{aligned}
S_{\mathcal{T}_{\alpha}^{\phantom{\alpha}}\!\!,v_{\beta}^{\phantom{\alpha}}}(\hbar\omega,\hat{\vn{M}})=&
\left[
S_{v_{\beta}^{\phantom{\alpha}},\mathcal{T}_{\alpha}^{\phantom{\alpha}}}(\hbar\omega,\hat{\vn{M}})
\right]^{*}
,
\\
{\rm Re}
[
S_{v_{\beta}^{\phantom{\alpha}},\mathcal{T}_{\alpha}^{\phantom{\alpha}}}(\hbar\omega,-\hat{\vn{M}})
]
=&
-{\rm Re}
[
S_{v_{\beta}^{\phantom{\alpha}},\mathcal{T}_{\alpha}^{\phantom{\alpha}}}(\hbar\omega,\hat{\vn{M}})
],\\
{\rm Im}
[
S_{v_{\beta}^{\phantom{\alpha}},\mathcal{T}_{\alpha}^{\phantom{\alpha}}}(\hbar\omega,-\hat{\vn{M}})
]
=&
{\rm Im}
[
S_{v_{\beta}^{\phantom{\alpha}},\mathcal{T}_{\alpha}^{\phantom{\alpha}}}(\hbar\omega,\hat{\vn{M}})
],
\end{aligned}
\ee
from which follows 
\bege
S_{\mathcal{T}_{\alpha}^{\phantom{\alpha}}\!\!,v_{\beta}^{\phantom{\alpha}}}(\hbar\omega,\hat{\vn{M}})=
-
S_{v_{\beta}^{\phantom{\alpha}},\mathcal{T}_{\alpha}^{\phantom{\alpha}}}(\hbar\omega,-\hat{\vn{M}})
\ee
and thus 
\bege
G^{\rm R}_{\mathcal{T}_{\alpha}^{\phantom{\alpha}}\!\!,v_{\beta}^{\phantom{\alpha}}}(\hbar\omega,\hat{\vn{M}})=
-
G^{\rm R}_{v_{\beta}^{\phantom{\alpha}},\mathcal{T}_{\alpha}^{\phantom{\alpha}}}(\hbar\omega,-\hat{\vn{M}}).
\ee
This identity allows us to rewrite
the magnetization-dynamics induced 
current density, Eq.~\eqref{eq_induced_current}, in 
terms of the torkance tensor as 
\bege\label{eq_jvol_from_torkance}
j_{\alpha}(t)=\frac{1}{V}
\sum_{\beta}
{\torkance}_{\beta\alpha}(-\hat{\vn{M}}(t))
\left(
\hat{\vn{M}}(t)
\times
\frac{d\hat{\vn{M}}(t)}{dt}
\right)_{\beta}.
\ee
Eq.~\eqref{eq_jvol_from_torkance} is the central result of this
subsection. It shows that it is very convenient 
to discuss the ISOT in terms
of the very same torkance tensor $\vn{\torkance}$ as the SOT. 
We note in passing that the torque-velocity
correlations, which the torkance measures, govern also the
Dzyaloshinskii-Moriya 
interaction~\cite{mothedmisot,phase_space_berry}.

It is convenient to decompose the torkance tensor 
into two components that are even and odd with 
respect to magnetization reversal, respectively~\cite{ibcsoit}:
$\vn{\torkance}(\hat{\vn{M}})
=\vn{\torkance}^{\rm even}(\hat{\vn{M}})
+\vn{\torkance}^{\rm odd}(\hat{\vn{M}})$, where
$\vn{\torkance}^{\rm even}(\hat{\vn{M}})=
[\vn{\torkance}(\hat{\vn{M}})
+\vn{\torkance}(-\hat{\vn{M}})]/2$ and
$\vn{\torkance}^{\rm odd}(\hat{\vn{M}})
=[\vn{\torkance}(\hat{\vn{M}})-
\vn{\torkance}(-\hat{\vn{M}})]/2$.
Separating 
$j_{\alpha}$ into the components 
due to $\vn{\torkance}^{\rm even}(\hat{\vn{M}})$ 
and $\vn{\torkance}^{\rm odd}(\hat{\vn{M}})$ yields
\bege
\begin{aligned}\label{eq_even_odd_current_density}
j_{\alpha}^{\rm even}(t)
\!&=\!
\frac{1}{V}
\!\sum_{\beta}\!
{\torkance}^{\rm even}_{\beta\alpha}(\hat{\vn{M}}(t))
\!\left(\!
\hat{\vn{M}}(t)
\!\times\!
\frac{d\hat{\vn{M}}(t)}{dt}
\!\right)_{\beta},\\
j_{\alpha}^{\rm odd}(t)
\!&=\!
-\frac{1}{V}
\!\sum_{\beta}\!
{\torkance}^{\rm odd}_{\beta\alpha}(\hat{\vn{M}}(t))
\!\left(\!
\hat{\vn{M}}(t)
\!\times\!
\frac{d\hat{\vn{M}}(t)}{dt}
\!\right)_{\beta}.
\end{aligned}
\ee

\subsection{Completing the response matrix}
\label{sec_completing_response_matrix}
When the electronic system is perturbed due to the time-dependence of
the exchange field direction a current density is induced according
to Eq.~\eqref{eq_induced_current}. This induced electric current is
not the only response of the electrons to this time dependent perturbation:
Additionally, the 
torque $-V\vn{\Lambda}(\hat{\vn{M}}\times\frac{d\,\hat{\vn{M}}}{d\,t})$
acts on the magnetization, where
\bege\label{eq_lambda_torque}
\Lambda_{\alpha\beta}=-
\frac{1}{V}\!
\lim_{\omega\to 0}\!
\frac{{\rm Im}G_{\mathcal{T}_{\alpha}^{\phantom{\alpha}},\mathcal{T}_{\beta}^{\phantom{\alpha}}}^{\rm R}
(\hbar\omega,\hat{\vn{M}})}{\hbar\omega}.
\ee
The sum of all torques on the magnetization has to be zero from the
point of view of an observer that rotates together with the magnetization:
\bege\label{eq_torque_balance}
0=\vn{\torkance}\vn{E}
-
V\vn{\Lambda}
\left(
\hat{\vn{M}}\times\frac{d\,\hat{\vn{M}}}{d\,t}
\right)
+\mu_{0}MV\hat{\vn{M}}\times\vn{H}^{\rm eff}.
\ee
Here, the first term on the right-hand side is the SOT. 
Torques such as the Gilbert damping torque, which are exerted on the
magnetization due to the magnetization dynamics, are described by 
the second term.
The third term summarizes torques due to external magnetic fields
and due to magnetic anisotropy. $M$ in the third term is the
magnetization, i.e., $MV$ is the magnetic moment.
In the presence of SOTs, the extended Landau-Lifshitz-Gilbert equation
runs
\bege\label{eq_llg}
\frac{d\,\hat{\vn{M}}}{d\,t}=
-|\gamma| \hat{\vn{M}}\times\vn{H}^{\rm eff}
+\vn{\alpha}\hat{\vn{M}}\times\frac{d\,\hat{\vn{M}}}{d\,t}
-\frac{|\gamma|\vn{\torkance}\vn{E}}{\mu_{0}MV},
\ee
where $\gamma=g\mu_{0}\mu_{\rm B}/\hbar$ is the gyromagnetic
ratio and $\vn{\alpha}$ is the Gilbert damping tensor.
Comparison of Eq.~\eqref{eq_llg} and Eq.~\eqref{eq_torque_balance}
leads to 
\bege\label{eq_gamma}
\frac{1}{\gamma}
=\frac{1}{2\mu_{0}M}
\sum_{\alpha\beta\delta}
\epsilon_{\alpha\beta\delta}
\Lambda_{\alpha\beta}^{\rm odd}\hat{M}_{\delta},
\ee
where $\epsilon_{\alpha\beta\delta}$ 
is the Levi-Civita symbol, and
\bege\label{eq_gilb_damp}
\vn{\alpha}=\frac{|\gamma| \vn{\Lambda}^{\rm even}_{\phantom{e}}}
{M\mu_{0}}.
\ee
It is straightforward to show that Eq.~\eqref{eq_gilb_damp} 
combined with Eq.~\eqref{eq_lambda_torque} reproduces 
the Gilbert damping expressions used within \textit{ab initio} 
calculations~\cite{damping_cpa_ebert}. 
In the absence of SOI it is found that~\cite{spin_dynamics_qian_vignale} 
\bege
\Lambda_{\alpha\beta}^{\rm odd}
=
-\frac{\hbar}{2\mu_{\rm B}}\sum_{\gamma}\epsilon_{\alpha\beta\gamma}M_{\gamma}.
\ee
Inserting this result into Eq.~\eqref{eq_gamma} leads to the 
expected nonrelativistic
value of $\gamma=-\frac{2\mu_{0}\mu_{\rm B}}{\hbar}$ and $g=-2$.

If we consider the coupled problem where both the
electric field and
the magnetization dynamics drive both the 
electric current and induce torques,
the even torkance $\vn{\torkance}^{\rm even}$ determines 
the off-diagonal elements of
the symmetric part $\vn{A}^{\rm s}$ of the corresponding 
linear response matrix, 
while the odd torkance $\vn{\torkance}^{\rm odd}$ determines 
those
of the antisymmetric part $\vn{A}^{\rm a}$:
\bege\label{eq_a_matrix_linear_response}
\begin{aligned}
\begin{pmatrix}
\vn{j}\\
\vn{T}/V
\end{pmatrix}=&
\left[
\vn{A}^{\rm s}(\hat{\vn{M}})+\vn{A}^{\rm a}(\hat{\vn{M}})
\right]
\begin{pmatrix}
\vn{E}\\
\hat{\vn{M}}\times \frac{d \hat{\vn{M}}}{dt}
\end{pmatrix}\\
\vn{A}^{\rm s}(\hat{\vn{M}})=&
\begin{pmatrix}
\vht{\sigma}^{\rm even}(\hat{\vn{M}}) &(\vn{\torkance}^{\rm even}(\hat{\vn{M}}))^{\rm T}/V\\
\vn{\torkance}^{\rm even}(\hat{\vn{M}})/V & -\vht{\Lambda}^{\rm even}(\hat{\vn{M}})
\end{pmatrix}\\
\vn{A}^{\rm a}(\hat{\vn{M}})=&
\begin{pmatrix}
\vht{\sigma}^{\rm odd}(\hat{\vn{M}}) &-(\vn{\torkance}^{\rm odd}(\hat{\vn{M}}))^{\rm T}/V\\
\vn{\torkance}^{\rm odd}(\hat{\vn{M}})/V & -\vht{\Lambda}^{\rm odd}(\hat{\vn{M}})
\end{pmatrix}.
\end{aligned}
\ee
Here, $\vn{\sigma}$ is the tensor of electrical conductivity. 
The torque $\vn{T}$ in the first equation, i.e., $\vn{T}=
\vn{\torkance}\vn{E}-
V\vn{\Lambda}\left(
\hat{\vn{M}}\times\frac{d\,\hat{\vn{M}}}{d\,t}
\right)$, is the torque on the magnetization due to the response of
the electrons to the two perturbations $\vn{E}$ 
and $\frac{d\,\hat{\vn{M}}}{d\,t}$. 
According to Eq.~\eqref{eq_torque_balance} the sum of this torque and
the torques due to magnetic anisotropy and external magnetic fields is zero. 
Due to the Onsager 
relation $\sigma_{\alpha\beta}(\hat{\vn{M}})=\sigma_{\beta\alpha}(-\hat{\vn{M}})$
the even part of the conductivity tensor is symmetric, i.e.,
$\sigma_{\alpha\beta}^{\rm even}(\hat{\vn{M}})=\sigma_{\beta\alpha}^{\rm even}(\hat{\vn{M}})$,
while the odd part 
is antisymmetric, i.e.,
$\sigma_{\alpha\beta}^{\rm odd}(\hat{\vn{M}})=-\sigma_{\beta\alpha}^{\rm odd}(\hat{\vn{M}})$~\cite{birss}. 
Similarly, $\Lambda_{\alpha\beta}^{\rm even}(\hat{\vn{M}})=\Lambda_{\beta\alpha}^{\rm even}(\hat{\vn{M}})$
and $\Lambda_{\alpha\beta}^{\rm odd}(\hat{\vn{M}})=-\Lambda_{\beta\alpha}^{\rm odd}(\hat{\vn{M}})$.
Consequently, $\vn{A}^{\rm s}(\hat{\vn{M}})$ is indeed 
symmetric and additionally even 
with respect to
magnetization reversal. Likewise, $\vn{A}^{\rm a}(\hat{\vn{M}})$ is 
indeed antisymmetric and
additionally odd with respect to magnetization reversal.
Therefore, the linear response 
matrix $\vn{A}(\hat{\vn{M}})=
\vn{A}^{\rm s}(\hat{\vn{M}})+\vn{A}^{\rm a}(\hat{\vn{M}})$ satisfies the symmetry
\bege\label{eq_onsager_a}
(\vn{A}(\hat{\vn{M}}))^{\rm T}=\vn{A}(-\hat{\vn{M}}),
\ee
which summarizes
the Onsager relations of $\vn{\sigma}$, $\vn{\Lambda}$ 
and $\vn{\torkance}$ in a compact form. 

Eq.~\eqref{eq_lambda_torque} and
Eq.~\eqref{eq_a_matrix_linear_response}
are the central results of this subsection.
They show that Gilbert damping
$\vn{\alpha}$ (Eq.~\eqref{eq_gilb_damp}), 
gyromagnetic ratio $\gamma$ (Eq.~\eqref{eq_gamma})
as well as ISOT (Eq.~\eqref{eq_induced_current}) can be
extracted coherently and consistently from time-dependent
perturbation theory, where the perturbation due to 
magnetization dynamics is given by Eq.~\eqref{eq_time_dependent_perturbation}. 

From the point of view
of adiabatic electron dynamics in a time-dependent 
Hamiltonian, Eq.~\eqref{eq_time_dependent_hamil}, it is natural
to consider the precession of the exchange field 
as perturbation. The electronic system responds to this perturbation
by the ISOT current, Eq.~\eqref{eq_induced_current}. Additionally, it
responds by 
the torque $\vn{T}=-V\vn{\Lambda}(\hat{\vn{M}}\times\frac{d\hat{\vn{M}}}{dt})$
described by Eq.~\eqref{eq_lambda_torque}.
However, when the Onsager reciprocity principle is used to
relate SOT and ISOT in a phenomenological approach
typically a different point of view is taken:
The effective magnetic field $\vn{H}^{\rm eff}$ is considered as a 
thermodynamic force and the time-derivative of magnetization 
plays the role of the associated thermodynamic 
flux~\cite{spin_motive_force_hals_brataas}.
Instead of considering the response 
of $(\vn{j},\vn{T}/V)^{\rm T}$ to the perturbation
$(\vn{E},\hat{\vn{M}}\times\frac{d\hat{\vn{M}}}{dt})^{\rm T}$
as we do in Eq.~\eqref{eq_a_matrix_linear_response} one
considers then instead the response of the thermodynamic fluxes
$(\frac{d \hat{\vn{M}} }{dt},\vn{j})^{\rm T}$
to the thermodynamic forces $(\vn{H}^{\rm eff},\vn{E})^{\rm T}$.
Interestingly, $\frac{d\hat{\vn{M}}}{dt}$ appears then as a
response rather than as a perturbation.
However, both formulations of the reciprocity between SOT and ISOT
are equivalent. 
\subsection{Many-electron response functions}
\label{sec_many_electron_resp}
In the previous two subsections we discussed SOT and ISOT based on the
effective single-particle Hamiltonian defined in
Eq.~\eqref{eq_time_dependent_hamil}, where the exchange field
$\Bxc^{\rm xc}(\vn{r})$ needs to be obtained self-consistently
within LSDA. In this subsection we
consider SOT and ISOT from the interacting many-electron point
of view. 

When a small static electric field $\vn{E}$ is applied to a magnet with
broken inversion symmetry its magnetization will assume a new
direction $\hat{\vn{M}}+\delta\hat{\vn{M}}$ 
due to the action of the SOT. We assume that $\vn{E}$ is
sufficiently small to ensure that the magnetization is not switched
and that 
$\hat{\vn{M}}+\delta\hat{\vn{M}}$ is time-independent.
Within linear response the relation between $\delta\hat{\vn{M}}$ 
and $\vn{E}$ is given by
\bege\label{eq_tilt_mag}
\delta\hat{\vn{M}}=\frac{1}{MV}\vn{\Xi}(\hat{\vn{M}})\vn{E}
\ee
with
\bege\label{eq_xi_spin_velocity_correlation}
\Xi_{\alpha\beta}(\hat{\vn{M}})=\lim_{\omega\to 0}\frac{e}{i\omega\hbar}
\mathcal{G}_{m_{\alpha}^{\phantom{\alpha}}\!,v_{\beta}^{\phantom{\alpha}}}^{\rm R}(\hbar\omega,\hat{\vn{M}}),
\ee
where
\bege\label{eq_spin_velocity_correlation}
\mathcal{G}_{m_{\alpha}^{\phantom{\alpha}},v_{\beta}^{\phantom{\alpha}}}^{\rm
  R}(\hbar\omega,\hat{\vn{M}})
=-i\int\limits_{0}^{\infty}dt e^{i\omega t}
\left\langle
[
m_{\alpha}^{\phantom{\alpha}} (t),v_{\beta}^{\phantom{\alpha}} (0)
]_{-}
\right\rangle
\ee
is the retarded spin-moment velocity correlation function. While
the correlation functions defined in
Eq.~\eqref{eq_velocity_torque_correlation},
Eq.~\eqref{eq_torque_velocity_correlation} and
Eq.~\eqref{eq_lambda_torque} are evaluated based on the Kohn-Sham
eigenfunctions of the 
effective single-particle Hamiltonian Eq.~\eqref{eq_static_Hamiltonian}, Eq.~\eqref{eq_spin_velocity_correlation}
has to be evaluated based on the interacting many-electron wave
functions of the system, i.e., 
\bege
\begin{aligned}
\mathcal{G}_{m_{\alpha}^{\phantom{\alpha}},v_{\beta}^{\phantom{\alpha}}}^{\rm
  R}(\hbar\omega,\hat{\vn{M}})
=&\sum_{n}
\hbar\Bigl [
\frac{
\langle\Psi_0|m_{\alpha}|\Psi_n \rangle
\langle \Psi_n|v_{\beta}|\Psi_0\rangle
}
{\mathcal{E}_0-\mathcal{E}_n+\hbar\omega+i\eta}\\
&-
\frac{
\langle\Psi_0|v_{\beta}|\Psi_n \rangle
\langle \Psi_n|m_{\alpha}|\Psi_0\rangle
}
{\mathcal{E}_n-\mathcal{E}_0+\hbar\omega+i\eta}
\Bigr],
\end{aligned}
\ee
where $\Psi_0$ is the ground state
and $\Psi_n$ with $n>0$ are the excited states.
The energies of the ground state and of the excited states are
$\mathcal{E}_0$ and $\mathcal{E}_n$, respectively. 
We use the symbol $\mathcal{G}^{\rm R}$ to denote the retarded 
many-electron response
functions while we use $G^{\rm R}$ to denote the 
retarded Kohn-Sham single-particle
response functions.

We can quantify the SOT that gives rise to the 
rotation of magnetization $\delta\hat{\vn{M}}$ in Eq.~\eqref{eq_tilt_mag}
in terms of the magnetic field $\vn{H}^{\rm SOT}$ 
that would need to be applied perpendicular to $\hat{\vn{M}}$
to achieve the 
same tilt $\delta\hat{\vn{M}}$ without applied electric field $\vn{E}$. 
The relation between $\delta\hat{\vn{M}}$ and
$\vn{H}^{\rm SOT}$ is described by the transverse magnetic 
susceptibility $\vn{\chi}$:
\bege\label{eq_tilt_mag_hsot}
M\delta\hat{\vn{M}}=\vn{\chi}(\hat{\vn{M}})\vn{H}^{\rm SOT},
\ee
where 
\bege\label{eq_transverse_mag_sus}
\chi_{\alpha\beta}(\hat{\vn{M}})
=
-\frac{\mu_0}{V\hbar}
\mathcal{G}_{m_{\alpha}^{\phantom{\alpha}},m_{\beta}^{\phantom{\alpha}}}^{\rm
  R}(\hbar\omega=0,\hat{\vn{M}}).
\ee
The static transverse magnetic 
susceptibility $\chi(\vn{\hat{M}})$ contains the information
on the magnetic anisotropy~\cite{damping_garate_macdonald}: 
When the magnetization
is tilted away from the easy axis due to 
the applied transverse magnetic field $\vn{H}^{\rm SOT}$, 
the additional internal magnetic
field 
\bege\label{eq_internal_mag_field}
\vn{H}^{\rm MAE}=-M[\vn{\chi}(\hat{\vn{M}})]^{-1}\delta\hat{\vn{M}}
\ee
due to magnetic anisotropy acts on the magnetization. 
The tilt $\delta\hat{\vn{M}}$ is such that $\vn{H}^{\rm MAE}+\vn{H}^{\rm SOT}=0$. 
Equating the right-hand sides of Eq.~\eqref{eq_tilt_mag} 
and Eq.~\eqref{eq_tilt_mag_hsot}
we obtain an expression for the magnetic field $\vn{H}^{\rm SOT}$:
\bege
\vn{H}^{\rm SOT}=
\frac{1}{V}
[\vn{\chi}(\hat{\vn{M}})]^{-1}
\vn{\Xi}(\hat{\vn{M}})\vn{E}.
\ee
This magnetic field exerts 
the torque $\mu_0 MV \hat{\vn{M}}\times \vn{H}^{\rm SOT}$ on the 
magnetization. Exactly the same torque acts on the magnetization when
the electric field $\vn{E}$ is applied instead of the magnetic field
$\vn{H}^{\rm SOT}$, i.e., the SOT is given 
by $\mu_0 MV \hat{\vn{M}}\times \vn{H}^{\rm SOT}$. 
The corresponding torkance can be written as
\bege\label{eq_torkance_mbf}
\tilde{\vn{\torkance}}(\hat{\vn{M}})
=\mu_0 M \hat{\vn{M}}\times 
[\vn{\chi}(\hat{\vn{M}})]^{-1}
\vn{\Xi}(\hat{\vn{M}}).
\ee
The applicability of Eq.~\eqref{eq_torkance_tensor}
is restricted to LSDA,
because it is based on the torque
operator $\vn{\mathcal{T}}$ and hence on the 
exchange field $\Bxc^{\rm xc}(\vn{r})$. In contrast,
Eq.~\eqref{eq_torkance_mbf} provides a general formulation of the torkance.
 
In order to show that Eq.~\eqref{eq_torkance_mbf} 
reduces to Eq.~\eqref{eq_torkance_tensor} within LSDA, i.e.,
$\tilde{\vn{\torkance}}(\hat{\vn{M}})
=\vn{\torkance}(\hat{\vn{M}})$,
we need to express the many-electron
response functions $\vn{\Xi}(\hat{\vn{M}})$ 
and $\vn{\chi}(\hat{\vn{M}})$ through the corresponding single-particle
Kohn-Sham response functions
\bege\label{eq_spin_veloc_ks}
\Xi^{\rm KS}_{\alpha\beta}(\hat{\vn{M}},\vn{r})
=\lim_{\omega\to 0}\frac{e}{i\omega\hbar}
G_{m_{\alpha}^{\phantom{\alpha}}(\vn{r}),v_{\beta}^{\phantom{\alpha}}}^{\rm R}(\hbar\omega,\hat{\vn{M}})
\ee
and
\bege
\chi^{\rm KS}_{\alpha\beta}(\hat{\vn{M}},\vn{r},\vn{r}')=
-\frac{\mu_0}{\hbar}
G_{
m_{\alpha}^{\phantom{\alpha}}(\vn{r}),
m_{\beta}^{\phantom{\alpha}}(\vn{r}')
}^{\rm R}(\hbar\omega=0,\hat{\vn{M}}),
\ee
where $m_{\alpha}^{\phantom{\alpha}}(\vn{r})$ is the operator of 
spin magnetic moment
density at position $\vn{r}$, i.e., 
$\int d^3 r\, m_{\alpha}^{\phantom{\alpha}}(\vn{r})=
m_{\alpha}^{\phantom{\alpha}}=-\mu_{\rm B}\sigma_{\alpha}^{\phantom{\alpha}}$.
When an electric field $\vn{E}$ is applied to the system the
transverse component of the change of magnetization 
at position $\vn{r}$, i.e., $m(\vn{r})\delta\hat{\vn{M}}(\vn{r})$, 
is described by the integral equation
\bege\label{eq_xi_integro}
\begin{aligned}
&m(\vn{r})\delta\hat{\vn{M}}(\vn{r})=
\vn{\Xi}^{\rm KS}(\hat{\vn{M}},\vn{r})\vn{E}+\\
&\,\,\,\,+\frac{1}{\mu_0}\int d^3 r'
\vn{\chi}^{\rm KS}(\hat{\vn{M}},\vn{r},\vn{r}')
\Bxc^{\rm xc}(\vn{r}')\delta\hat{\vn{M}}(\vn{r}').
\end{aligned}
\ee
The second term on the right hand side takes into account that
within LSDA the quasiparticles respond not only to the applied fields
but also to the induced fields. In order to solve this integral
equation approximatively, we assume that the change of magnetization 
direction is independent of 
position, i.e., $\delta\hat{\vn{M}}(\vn{r})=\delta\hat{\vn{M}}$. Multiplying
both sides of Eq.~\eqref{eq_xi_integro} 
by $\Bxc^{\rm xc}(\vn{r})\hat{\vn{M}}\times$ from the left, 
and integrating over position $\vn{r}$
we obtain
\bege\label{eq_integrated_ks}
\begin{aligned}
&\bar{\Omega}^{\rm xc}MV(\hat{\vn{M}}\times\delta\hat{\vn{M}})
=\vn{\torkance}(\hat{\vn{M}})\vn{E}\\
&-
\frac{1}{\hbar}\sum_{\alpha\beta}
\hat{\vn{e}}_{\alpha}
G^{\rm R}_{\mathcal{T}_{\alpha}\mathcal{T}_{\beta}}
(\hbar\omega=0,\hat{\vn{M}})
[\hat{\vn{M}}\times \delta\hat{\vn{M}}]_{\beta}.
\end{aligned}
\ee
The average exchange field on the left-hand side is defined as
\bege
\bar{\Omega}^{\rm xc}
=
\frac{
\int d^3 r\,
\Bxc^{\rm xc}(\vn{r})
m(\vn{r})
}
{
\int d^3 r'
m(\vn{r}')
}
=
\frac{
\int d^3 r\,
\Bxc^{\rm xc}(\vn{r})
m(\vn{r})
}
{
MV
}.
\ee
To obtain the first term on the right-hand side 
of Eq.~\eqref{eq_integrated_ks} 
we made use of
\bege
\hat{\vn{M}}\!\times\!\int d^3\,r\,\vn{\Xi}^{\rm KS}(\hat{\vn{M}},\vn{r})
\Bxc^{\rm xc}(\vn{r})=\vn{\torkance},
\ee
which follows from comparison of Eq.~\eqref{eq_torkance_tensor} 
and Eq.~\eqref{eq_spin_veloc_ks}.
Solving Eq.~\eqref{eq_integrated_ks} for $\delta\hat{\vn{M}}$ and 
comparing to Eq.~\eqref{eq_tilt_mag} yields the following expression for
$\vn{\Xi}(\hat{\vn{M}})$:
\bege\label{eq_xi_final}
\vn{\Xi}(\hat{\vn{M}})=-\hat{\vn{M}}\!\times\!
\left[
\bar{\Omega}^{\rm xc}
\!+\!\frac{
G^{\rm R}_{\vn{\mathcal{T}}\vn{\mathcal{T}}}
(\hbar\omega=0,\hat{\vn{M}})
}
{MV\hbar}
\right]^{-1}\!\!\!\!\!\vn{\torkance}(\hat{\vn{M}}).
\ee

In order to obtain an expression for $\vn{\chi}(\hat{\vn{M}})$ 
in Eq.~\eqref{eq_tilt_mag_hsot} we need to 
replace $\vn{\Xi}^{\rm KS}(\hat{\vn{M}},\vn{r})\vn{E}$ 
in Eq.~\eqref{eq_xi_integro} 
by $\int d^3\,r'\vn{\chi}^{\rm KS}(\hat{\vn{M}},\vn{r},\vn{r}')\vn{H}^{\rm SOT}$, 
which yields the equation
\begin{gather}\label{eq_chi_integro}
\begin{aligned}
&m(\vn{r})\delta\hat{\vn{M}}(\vn{r})=
\int d^3\,r'\vn{\chi}^{\rm KS}(\hat{\vn{M}},\vn{r},\vn{r}')
\Bxc^{\rm xc}(\vn{r}')\frac{\vn{H}^{\rm SOT}}{\bar{\Omega}^{\rm xc}}
+\\
&\,\,\,\,+\frac{1}{\mu_0}\int d^3 r'
\vn{\chi}^{\rm KS}(\hat{\vn{M}},\vn{r},\vn{r}')
\Bxc^{\rm xc}(\vn{r}')\delta\hat{\vn{M}}(\vn{r}'),
\end{aligned}\raisetag{1.2\baselineskip}
\end{gather}
where we replaced the 
magnetic field $\vn{H}^{\rm SOT}$ 
by $\vn{H}^{\rm SOT}\Bxc^{\rm xc}(\vn{r}')/\bar{\Omega}^{\rm xc}$,
because both magnetic fields produce the same torque 
on the magnetization~\cite{damping_garate_macdonald}:
\bege
\frac{\mu_0 \hat{\vn{M}}\!\! \times\!\!\vn{H}^{\rm SOT}}{\bar{\Omega}^{\rm xc}}
\!\int\! d^3 r\,m(\vn{r})\Bxc^{\rm xc}(\vn{r})
=\mu_0V\vn{M}\!\times\!\vn{H}^{\rm SOT}.
\ee
Multiplying
both sides of Eq.~\eqref{eq_chi_integro}
by $\Bxc^{\rm xc}(\vn{r})\hat{\vn{M}}\times$ from the left,
and integrating over position $\vn{r}$,
we obtain
\bege\label{eq_integrated_chi_ks}
\begin{aligned}
&\bar{\Omega}^{\rm xc}MV(\hat{\vn{M}}\times\delta\hat{\vn{M}})
=\\
&-
\frac{\mu_0}{\hbar\bar{\Omega}^{\rm xc}}\sum_{\alpha\beta}
\hat{\vn{e}}_{\alpha}
G^{\rm R}_{\mathcal{T}_{\alpha}\mathcal{T}_{\beta}}
(\hbar\omega=0,\hat{\vn{M}})
[\hat{\vn{M}}\times \vn{H}^{\rm SOT}]_{\beta}\\
&-
\frac{1}{\hbar}\sum_{\alpha\beta}
\hat{\vn{e}}_{\alpha}
G^{\rm R}_{\mathcal{T}_{\alpha}\mathcal{T}_{\beta}}
(\hbar\omega=0,\hat{\vn{M}})
[\hat{\vn{M}}\times \delta\hat{\vn{M}}]_{\beta}.
\end{aligned}
\ee
Comparing Eq.~\eqref{eq_integrated_chi_ks} and Eq.~\eqref{eq_integrated_ks}
leads to
\bege\label{eq_ve_mh}
\vn{\torkance}(\hat{\vn{M}})\vn{E}
\!=\!
-
\frac{\mu_0}{\hbar\bar{\Omega}^{\rm xc}}
G^{\rm R}_{\vn{\mathcal{T}}\vn{\mathcal{T}}}
(\hbar\omega\!=\!0,\hat{\vn{M}})
\!\left[
\hat{\vn{M}}\!\times\! \vn{H}^{\rm SOT}
\right].
\ee
In the absence of SOI, $G^{\rm R}_{\vn{\mathcal{T}}\vn{\mathcal{T}}}
(\hbar\omega\!=\!0,\hat{\vn{M}})$ is 
given by (see Appendix~\ref{app_mae})
\bege\label{eq_gtt}
G^{\rm R}_{\vn{\mathcal{T}}\vn{\mathcal{T}}}
(\hbar\omega\!=\!0,\hat{\vn{M}})=
-\hbar M V \bar{\Omega}^{\rm xc}
\left[
1-\hat{\vn{M}}^{\rm T}\hat{\vn{M}}
\right].
\ee
We assume that the magnetic anisotropy is small compared to the 
exchange splitting. In this case we can 
approximate $G^{\rm R}_{\vn{\mathcal{T}}\vn{\mathcal{T}}}
(\hbar\omega\!=\!0,\hat{\vn{M}})$ in Eq.~\eqref{eq_ve_mh} 
by Eq.~\eqref{eq_gtt} and obtain
\bege\label{eq_te_mh_fi}
\vn{T}^{\rm SOT}=
\mu_0 M V \left[
\hat{\vn{M}}\!\times\! \vn{H}^{\rm SOT}
\right]
=\vn{\torkance}(\hat{\vn{M}})\vn{E}
.
\ee
Eq.~\eqref{eq_te_mh_fi} shows that the description of the SOT 
through Eq.~\eqref{eq_torkance_mbf} in terms of
many-electron response 
functions Eq.~\eqref{eq_xi_spin_velocity_correlation} 
and Eq.~\eqref{eq_transverse_mag_sus}
recovers
the single-particle expression Eq.~\eqref{eq_torkance_tensor}.

Solving Eq.~\eqref{eq_integrated_chi_ks} for $\delta\hat{\vn{M}}$
and comparing to Eq.~\eqref{eq_tilt_mag_hsot} yields 
\bege\label{eq_chi}
\begin{aligned}
&\vn{\chi}(\hat{\vn{M}})
=
\frac{\mu_0}{\hbar V\bar{\Omega}^{\rm xc}}
\hat{\vn{M}}\times
\Bigl[
\bar{\Omega}^{\rm xc}
+\\
&+\frac{
G^{\rm R}_{\vn{\mathcal{T}}\vn{\mathcal{T}}}
(\hbar\omega=0,\hat{\vn{M}})
}
{MV\hbar}
\Bigr]^{-1}
G^{\rm R}_{\vn{\mathcal{T}}\vn{\mathcal{T}}}
(\hbar\omega=0,\hat{\vn{M}})\hat{\vn{M}}\times,
\end{aligned}
\ee
where $\hat{\vn{M}}\times$ is a shorthand for the matrix
\bege\label{eq_cross_op}
\frac{1}{M}
\begin{pmatrix}
0 &-M_3 &M_2\\
M_3 &0 &-M_1\\
-M_2 &M_1 &0
\end{pmatrix}
=\hat{\vn{M}}\times.
\ee
Assuming that the anisotropy energy is much smaller than the
exchange splitting, we can approximate 
the rightmost $G^{\rm R}_{\vn{\mathcal{T}}\vn{\mathcal{T}}}$
in Eq.~\eqref{eq_chi} by Eq.~\eqref{eq_gtt} and obtain
\bege\label{eq_chi_final}
\vn{\chi}(\hat{\vn{M}})\!=\!-\mu_0 M \hat{\vn{M}}\!\times\!
\left[
\bar{\Omega}^{\rm xc}\!+\!\frac{
G^{\rm R}_{\vn{\mathcal{T}}\vn{\mathcal{T}}}
(\hbar\omega=0,\hat{\vn{M}})
}
{MV\hbar}
\right]^{-1}\!\!\!\!\!\!
\hat{\vn{M}}\times.
\ee
The difference between the right-hand side and the left-hand side of
Eq.~\eqref{eq_gtt} describes the magnetic 
anisotropy (see Appendix~\ref{app_mae}). 
Therefore, the remaining $G^{\rm R}_{\vn{\mathcal{T}}\vn{\mathcal{T}}}$ in
Eq.~\eqref{eq_chi_final} cannot be approximated by Eq.~\eqref{eq_gtt}.  
Inserting Eq.~\eqref{eq_chi_final} and Eq.~\eqref{eq_xi_final} into
Eq.~\eqref{eq_torkance_mbf} leads to the 
identity $\tilde{\vn{\torkance}}(\hat{\vn{M}})
=\vn{\torkance}(\hat{\vn{M}})$, showing again the equivalence between the
single-particle and the many-electron expressions, 
Eq.~\eqref{eq_torkance_tensor}
and Eq.~\eqref{eq_torkance_mbf}, respectively. 

Using Eq.~\eqref{eq_chi_final} we can rewrite Eq.~\eqref{eq_xi_final} as
\bege
\vn{\Xi}(\hat{\vn{M}})
=-\frac{1}{\mu_0 M}
\vn{\chi}(\hat{\vn{M}})
\hat{\vn{M}}\times\vn{\torkance}(\hat{\vn{M}}).
\ee
In this expression, $\hat{\vn{M}}\times\vn{\torkance}(\hat{\vn{M}})$ on the
right-hand side can be
interpreted in terms of a current-induced effective magnetic 
field $\vn{H}^{\rm SOT}
=-[\hat{\vn{M}}\times\vn{\torkance}(\hat{\vn{M}})\vn{E}]/(MV\mu_0)$.
The transverse magnetic susceptibility $\vn{\chi}(\hat{\vn{M}})$ describes
the response of the magnetization to $\vn{H}^{\rm SOT}$.

Next, we consider the generation of a current density $\vn{j}$ due to
a time-dependent applied magnetic 
field $\vn{H}^{\rm ext}(\omega,t)=\vn{H}^{\rm ext}(\omega)e^{-i\omega t}$.
Denoting the corresponding linear response tensor
by $\vn{\Phi}(\hat{\vn{M}},\omega)$ we can write
\bege\label{eq_isot_mb_phi_define}
\begin{aligned}
\vn{j}
&=\vn{\Phi}(\hat{\vn{M}},\omega)
\vn{H}^{\rm ext}(\omega)e^{-i\omega t}\\
&\simeq
\left.\frac{d\,\vn{\Phi}(\hat{\vn{M}},\omega)}{d\,\omega}\right|_{\omega=0}\!\!\!\omega \vn{H}^{\rm ext}(\omega)e^{-i\omega t}\\
&=i \vn{\Phi}' (\hat{\vn{M}})
\frac{d\,\vn{H}^{\rm ext}(\omega,t)}{d\,t},
\end{aligned}
\ee
where $\vn{\Phi}'(\hat{\vn{M}})$ denotes the frequency derivative, 
i.e., $\vn{\Phi}'(\hat{\vn{M}})=\left.\frac{d\,\vn{\Phi}(\hat{\vn{M}},\omega)}{d\,\omega}\right|_{\omega=0}$.
We used that $\vn{\Phi}(\hat{\vn{M}},\omega=0)$ does 
not generate an ISOT current and
we expanded $\vn{\Phi}(\hat{\vn{M}},\omega)$ up to first order in frequency.
Assuming that the field $\vn{H}^{\rm ext}(\omega,t)$ is transverse to 
magnetization, we can use the transverse magnetic 
susceptibility $\vn{\chi}(\hat{\vn{M}})$, Eq.~\eqref{eq_tilt_mag_hsot}, to
express it in terms of the 
corresponding tilt of the magnetization direction. This allows us to
relate $\vn{j}$ to the time-derivative of the magnetization direction:
\bege\label{eq_isot_mbf}
\begin{aligned}
\vn{j}&=i M\vn{\Phi}'(\hat{\vn{M}})[\vn{\chi}(\hat{\vn{M}})]^{-1}
\frac{d\,\hat{\vn{M}}}{d\,t}\\
&=-i M\vn{\Phi}'(\hat{\vn{M}})[\vn{\chi}(\hat{\vn{M}})]^{-1}
\hat{\vn{M}}\times
\left[\hat{\vn{M}}\times
\frac{d\,\hat{\vn{M}}}{d\,t}
\right].
\end{aligned}
\ee

We can use the retarded velocity spin-moment correlation function
\bege\label{eq_velocity_spin_correlation}
\mathcal{G}_{
v_{\alpha}^{\phantom{\alpha}},
m_{\beta}^{\phantom{\alpha}}
}^{\rm
  R}(\hbar\omega,\hat{\vn{M}})
=-i\int\limits_{0}^{\infty}dt e^{i\omega t}
\left\langle
[
v_{\alpha}^{\phantom{\alpha}} (t),
m_{\beta}^{\phantom{\alpha}} (0)
]_{-}
\right\rangle
\ee
to express $\vn{\Phi}'(\hat{\vn{M}})$ as follows:
\bege\label{eq_phi_velocity_spin_correlation}
\Phi'_{\alpha\beta}(\hat{\vn{M}})=
\lim_{\omega\to 0}\frac{d}{d\,\omega}\frac{e\mu_{0}}{\hbar V}
\mathcal{G}_{
v_{\alpha}^{\phantom{\alpha}},
m_{\beta}^{\phantom{\alpha}}
}^{\rm R}(\hbar\omega,\hat{\vn{M}})
\ee
The spectral densities of the Green functions defined in
Eq.~\eqref{eq_spin_velocity_correlation} and
in Eq.~\eqref{eq_velocity_spin_correlation} are given by
\bege
\begin{aligned}
S_{m_{\alpha}^{\phantom{\alpha}},v_{\beta}^{\phantom{\alpha}}} (t,t',\hat{\vn{M}})
=&
\frac{1}{2\pi}
\left\langle
[
m_{\alpha}(t)
,
v_{\beta}(t')
]_{-}\right\rangle,\\
S_{v_{\alpha}^{\phantom{\alpha}},m_{\beta}^{\phantom{\alpha}}}(t,t',\hat{\vn{M}})
=&
\frac{1}{2\pi}
\left\langle
[
v_{\alpha}(t)
,
m_{\beta}(t')
]_{-}
\right\rangle
\end{aligned}
\ee
and their Fourier transforms satisfy the relations
\bege
\begin{aligned}
S_{m_{\alpha}^{\phantom{\alpha}},v_{\beta}^{\phantom{\alpha}}}(\hbar\omega,\hat{\vn{M}})=&
\left[
S_{v_{\beta}^{\phantom{\alpha}},m_{\alpha}^{\phantom{\alpha}}}(\hbar\omega,\hat{\vn{M}})
\right]^{*}
\!,
\\
{\rm Re}
[
S_{v_{\beta}^{\phantom{\alpha}},m_{\alpha}^{\phantom{\alpha}}}(\hbar\omega,-\hat{\vn{M}})
]
=&
\phantom{-}{\rm Re}
[
S_{v_{\beta}^{\phantom{\alpha}},m_{\alpha}^{\phantom{\alpha}}}(\hbar\omega,\hat{\vn{M}})
],\\
{\rm Im}
[
S_{v_{\beta}^{\phantom{\alpha}},m_{\alpha}^{\phantom{\alpha}}}(\hbar\omega,-\hat{\vn{M}})
]
=&-
{\rm Im}
[
S_{v_{\beta}^{\phantom{\alpha}},m_{\alpha}^{\phantom{\alpha}}}(\hbar\omega,\hat{\vn{M}})
],
\end{aligned}
\ee
from which follows
\bege
S_{m_{\alpha}^{\phantom{\alpha}},v_{\beta}^{\phantom{\alpha}}}(\hbar\omega,\hat{\vn{M}})=
S_{v_{\beta}^{\phantom{\alpha}},m_{\alpha}^{\phantom{\alpha}}}(\hbar\omega,-\hat{\vn{M}})
\ee
and thus
\bege
\mathcal{G}^{\rm R}_{m_{\alpha}^{\phantom{\alpha}},v_{\beta}^{\phantom{\alpha}}}(\hbar\omega,\hat{\vn{M}})=
\mathcal{G}^{\rm R}_{v_{\beta}^{\phantom{\alpha}},m_{\alpha}^{\phantom{\alpha}}}(\hbar\omega,-\hat{\vn{M}})
\ee
and
\bege
\vn{\Phi}'(\hat{\vn{M}})=
\lim_{\omega\to 0}\frac{d}{d\,\omega}\frac{e\mu_{0}}{\hbar V}
\left[
\mathcal{G}_{
\vn{m},
\vn{v}
}^{\rm R}(\hbar\omega,-\hat{\vn{M}})
\right]^{\rm T}.
\ee
Using
Eq.~\eqref{eq_torkance_mbf}, 
$[\hat{\vn{M}}\times]^{\rm T}=-\hat{\vn{M}}\times$ 
(see Eq.~\eqref{eq_cross_op})
and the Onsager 
relation $\vn{\chi}(\hat{\vn{M}})=[\vn{\chi}(-\hat{\vn{M}})]^{\rm T}$
we can relate $\vn{\Phi}'(\hat{\vn{M}})$ and 
the torkance $\tilde{\vn{\torkance}}(\hat{\vn{M}})$
as follows:
\bege
\left[
\tilde{\vn{\torkance}}(-\hat{\vn{M}})
\right]^{\rm T}
=-iVM
\vn{\Phi}'(\hat{\vn{M}})
[\vn{\chi}(\hat{\vn{M}})]^{-1}
\hat{\vn{M}}\times.
\ee
This allows us to rewrite Eq.~\eqref{eq_isot_mbf} as
\bege
\begin{aligned}
\vn{j}=\frac{1}{V}\left[
\tilde{\vn{\torkance}}(-\hat{\vn{M}})
\right]^{\rm T}
\hat{\vn{M}}
\times
\frac{d\,\vn{M}}{d\,t}
\end{aligned}
\ee
in agreement with Eq.~\eqref{eq_jvol_from_torkance} derived 
earlier in the single-particle
formalism. 

The central result of this subsection is Eq.~\eqref{eq_torkance_mbf}, which
provides a general definition of the torkance that is not limited to the framework of 
Kohn-Sham theory. The reciprocity between direct and inverse SOT as
discussed in the previous subsection based on Kohn-Sham theory
remains valid within the many-electron response function formalism used in this subsection. 
\section{SOT and ISOT in bilayer systems}
\label{sec_angular_dependence_sot_isot}
In the following we discuss SOT and ISOT 
in magnetic bilayer systems
composed of a ferromagnetic layer (FM) deposited on a normal
metal (NM). 
When the electric field $\vn{E}={\rm E}_{x}\hat{\vn{e}}_{x}$ is 
applied in-plane along 
$x$ direction, 
the 
torques
satisfy
\begin{gather}\label{eq_teven}
\begin{aligned}
\vn{T}^{\rm even}(\hat{\vn{M}})\!=\!
{\rm E}_{x}
\hat{\vn{M}}
\!\!\times\!\!
(\!
\hat{\vn{e}}_{y}
\!\!\times\!\!
\hat{\vn{M}}\!
)
&[
A_{0}
\!+\!
A_{2}(\hat{\vn{e}}_{z}
\!\!\times\!\!
\hat{\vn{M}})^{2}
\!+\dots
]+\phantom{abc}\\
+{\rm E}_{x}
(\!
\hat{\vn{M}}
\!\!\times\!
\hat{\vn{e}}_{z}\!
)
(\!
\hat{\vn{M}}
\!\cdot\!
\hat{\vn{e}}_{x}\!
)
&[
B_{2}
\!+\!
B_{4}(\hat{\vn{e}}_{z}
\!\!\times\!\!
\hat{\vn{M}})^2+\!\dots
]
\end{aligned}\raisetag{1.2\baselineskip}
\end{gather}
and
\begin{gather}\label{eq_todd}
\begin{aligned}
\vn{T}^{\rm odd}(\hat{\vn{M}})\!=\!
{\rm E}_{x}
(\!
\hat{\vn{e}}_{y}
\!\!\times\!\!
\hat{\vn{M}}
\!)
[
C_{0}
\!+\!
C_{2}
(
\hat{\vn{e}}_{z}
\!\!\times\!\!
\hat{\vn{M}}
)^{2}
\!+\dots
]+&\phantom{abcdefgh}\\
+{\rm E}_{x}\hat{\vn{M}}
\!\!\times\!\!
(
\!\hat{\vn{M}}
\!\!\times\!
\hat{\vn{e}}_{z}
\!)
(\!
\hat{\vn{M}}
\!\cdot\!
\hat{\vn{e}}_{x}
\!)
[
D_{2}
\!+\!
D_{4}(\vn{e}_{z}
\!\!\times\!\!
\hat{\vn{M}})^2
\!+&\dots
]
\end{aligned}\raisetag{1.2\baselineskip}
\end{gather}
in bilayer systems composed of polycrystalline, 
disordered or amorphous layers with continuous rotational
symmetry around the $z$ 
axis~\cite{symmetry_spin_orbit_torques}.

To describe the ISOT in bilayer systems 
we consider instead of the current density $j_{\alpha}$ 
the current per
length 
$J_{\alpha}$, which is obtained by replacing
the current density operator $-ev_{\alpha}/V$ by
$-ev_{\alpha}/A$,
where $A$ is the cross sectional
area of the unit cell of the bilayer normal to the
stacking direction:
\bege
J_{\alpha}(t)=
\frac{1}{A}
\sum_{\beta}{\torkance}_{\beta\alpha}(-\hat{\vn{M}}(t))
\left(
\hat{\vn{M}}(t)
\times
\frac{d\hat{\vn{M}}(t)}{dt}
\right)_{\beta}.
\ee
Since the atom-resolved current is expected to
vary significantly between atomic layers in
bilayer systems, 
$J_{\alpha}$ is a suitable
definition of current density in such systems.
In terms of $\vn{J}$, the electric current flowing 
in $x$ direction is
given by $I_{x}=J_x L_y$, where $L_y$ is the length of the
system in $y$ direction, and 
similarly $I_{y}=J_y L_x$ is the
electric current in $y$ direction.
Separating 
$J_{\alpha}$ into the components 
due to $\vn{\torkance}^{\rm even}(\hat{\vn{M}})$ 
and $\vn{\torkance}^{\rm odd}(\hat{\vn{M}})$ yields
\bege
\begin{aligned}\label{eq_even_odd_currents}
J_{\alpha}^{\rm even}(t)
\!&=\!
\frac{1}{A}
\!\sum_{\beta}\!
{\torkance}^{\rm even}_{\beta\alpha}(\hat{\vn{M}}(t))
\!\left(\!
\hat{\vn{M}}(t)
\!\times\!
\frac{d\hat{\vn{M}}(t)}{dt}
\!\right)_{\beta},\\
J_{\alpha}^{\rm odd}(t)
\!&=\!
-\frac{1}{A}
\!\sum_{\beta}\!
{\torkance}^{\rm odd}_{\beta\alpha}(\hat{\vn{M}}(t))
\!\left(\!
\hat{\vn{M}}(t)
\!\times\!
\frac{d\hat{\vn{M}}(t)}{dt}
\!\right)_{\beta}.
\end{aligned}
\ee

In the following we discuss the  
magnetization-dynamics induced
current density 
$J_{x}$
in $x$ direction.
Using Eq.~\eqref{eq_teven} and Eq.~\eqref{eq_todd} 
in Eq.~\eqref{eq_even_odd_currents} we obtain  
\begin{gather}\label{eq_induced_even_current_x}
\begin{aligned}
J_{x}^{\,\rm even}(t)&=
\frac{A_0}{A}
\left[
\hat{\vn{M}}
\!\!\times\!\!
\left(\!
\hat{\vn{e}}_{y}
\!\!\times\!\!
\hat{\vn{M}}\!
\right)
\right]
\!\cdot\!
\left[
\hat{\vn{M}}
\!\!\times\!\!
\frac{d\hat{\vn{M}}}{dt}
\right]+\\
&\!\!\!\!\!\!\!\!+
\frac{A_2}{A}
\left[
\hat{\vn{M}}
\!\!\times\!\!
\left(\!
\hat{\vn{e}}_{y}
\!\!\times\!\!
\hat{\vn{M}}\!
\right)
\right]
\!\cdot\!
\left[
\hat{\vn{M}}
\!\!\times\!\!
\frac{d\hat{\vn{M}}}{dt}
\right]\!\!
\left(\!
\hat{\vn{e}}_{z}
\!\!\times\!\!
\hat{\vn{M}}\!
\right)^{2}\!+\\
&\!\!\!\!\!\!\!\!+\frac{B_2}{A}
\left(\!
\hat{\vn{M}}
\!\!\times\!\!
\hat{\vn{e}}_{z}\!
\right)
\!\cdot\!
\left[
\hat{\vn{M}}
\!\!\times\!\!
\frac{d\hat{\vn{M}}}{dt}
\right]\!
\left(\!
\hat{\vn{M}}
\!\cdot\!
\hat{\vn{e}}_{x}\!
\right)+\\
&\!\!\!\!\!\!\!\!+\frac{B_4}{A}
\left(\!
\hat{\vn{M}}
\!\!\times\!\!
\hat{\vn{e}}_{z}\!
\right)
\!\cdot\!
\left[
\hat{\vn{M}}
\!\!\times\!\!
\frac{d\hat{\vn{M}}}{dt}
\right]\!
\left(\!
\hat{\vn{M}}
\!\cdot\!
\hat{\vn{e}}_{x}\!
\right)\!\!
\left(\!
\hat{\vn{e}}_{z}
\!\!\times\!\!
\hat{\vn{M}}\!
\right)^2+\\
&\!\!\!\!\!\!\!\!+\cdots
\end{aligned}\raisetag{6\baselineskip}
\end{gather}
and
\begin{gather}\label{eq_induced_odd_current_x} 
\begin{aligned}
J_{x}^{\,\rm odd}(t)&=
-\frac{C_0}{A}\!
\left(\!
\hat{\vn{e}}_{y}
\!\!\times\!\!
\hat{\vn{M}}\!
\right)
\!\cdot\!
\left[\!
\hat{\vn{M}}
\!\!\times\!\!
\frac{d\hat{\vn{M}}}{dt}\!
\right]\\
&\!\!\!\!\!\!\!\!
-\frac{C_2}{A}
\left(\!
\hat{\vn{e}}_{y}
\!\!\times\!\!
\hat{\vn{M}}\!
\right)
\!\cdot\!
\left[
\hat{\vn{M}}
\!\!\times\!\!
\frac{d\hat{\vn{M}}}{dt}
\right]\!\!
\left(\!
\hat{\vn{e}}_{z}
\!\!\times\!\!
\hat{\vn{M}}\!
\right)^{2}\\
&\!\!\!\!\!\!\!\!-\frac{D_2}{A}
\left[
\hat{\vn{M}}
\!\!\times\!\!
\left(\!
\hat{\vn{M}}
\!\!\times\!\!
\hat{\vn{e}}_{z}\!
\right)
\right]
\!\!\cdot\!\!
\left[
\hat{\vn{M}}
\!\!\times\!\!
\frac{d\hat{\vn{M}}}{dt}
\right]\!\!
\left(\!
\hat{\vn{M}}
\!\cdot\!
\hat{\vn{e}}_{x}
\right)\\
&\!\!\!\!\!\!\!\!-\frac{D_4}{A}
\left[
\hat{\vn{M}}
\!\!\times\!\!
\left(\!
\hat{\vn{M}}
\!\!\times\!\!
\hat{\vn{e}}_{z}\!
\right)
\right]
\!\!\cdot\!\!
\left[\!
\hat{\vn{M}}
\!\!\times\!\!
\frac{d\hat{\vn{M}}}{dt}\!
\right]\!\!
\left(\!
\hat{\vn{M}}
\!\cdot\!
\hat{\vn{e}}_{x}\!
\right)\!\!
\left(\!
\hat{\vn{e}}_{z}
\!\!\times\!\!
\hat{\vn{M}}\!
\right)^2\\
&\!\!\!\!\!\!\!\!-\cdots.
\end{aligned}\raisetag{6\baselineskip}
\end{gather}

\subsection{Current densities induced by FMR through the inverse SOT}
First, we consider the case of FMR-driven magnetization 
precession around the $z$ axis 
in a circular orbit, i.e., 
\bege\label{eq_precession_around_z}
\hat{\vn{M}}(t)
=
\left[
\sin(\theta)\cos(\omega t),\sin(\theta)\sin(\omega t),\cos(\theta)
\right]^{\rm T},
\ee
where $\theta$ is the cone angle.
Inserting Eq.~\eqref{eq_precession_around_z} into 
Eqs.~\eqref{eq_induced_even_current_x} 
and~\eqref{eq_induced_odd_current_x} we obtain
\begin{gather}
\begin{aligned}\label{eq_induced_currents_z}
J_{x}^{\,\rm even}(t)\!&=\!
-\frac{\omega}{A}
\!\sin(\theta)\!\cos(\theta)\!\sin(\omega t)
\![
A_0
\!+\!
A_2\sin^2(\theta)
\!+\!
\cdots
],\\
J_{x}^{\,\rm odd}(t)\!&=\!
\frac{\omega}{A}
\!\sin(\theta)\!\cos(\omega t)\![C_0\!+\!C_2\sin^2(\theta)\!+\!\cdots]+\\
&\!\!\!\!\!\!+\frac{\omega}{A}
\!\sin(\theta)\!\cos(\omega t)
\![D_2\sin^2(\theta)\!+\!D_4\sin^4(\theta)\!+\!\cdots].
\end{aligned}\raisetag{2.6\baselineskip}
\end{gather}
For small cone angles $\theta$ the $\sin^2(\theta)$ factors suppress
the contributions from $A_2$, $C_2$, $D_2$ and further 
higher-order terms. In the small cone limit the ISOT for
magnetization precession around the $z$ axis can thus
be expressed in terms of the torkance for magnetization along
$z$, if $A_0={\torkance}^{\rm even}_{yx}(\hat{\vn{M}}=\hat{\vn{e}}_{z})$
and $C_0={\torkance}^{\rm odd}_{xx}(\hat{\vn{M}}=\hat{\vn{e}}_{z})$
are used.
Experiments~\cite{symmetry_spin_orbit_torques,
layer_thickness_dependence_current_induced_effective_field_TaCoFeB_Hayashi} 
and \textit{ab initio} calculations~\cite{ibcsoit} 
have found that $A_0$ and $C_0$ can
be of the same order of magnitude in AlO$_{x}$/Co/Pt and MgO/CoFeB/Ta.
The two contributions $J_{x}^{\,\rm even}(t)$ 
and $J_{x}^{\,\rm odd}(t)$ 
are therefore expected to exhibit similar amplitudes. 
Since $J_{x}^{\,\rm even}(t)\propto \sin(\omega t)$ 
while $J_{x}^{\,\rm odd}(t)\propto \cos(\omega t)$
the even and odd part are phase-shifted with respect to each other.

Next, we consider FMR-driven magnetization 
precession around the $y$ axis. 
In this case the magnetization
follows an elliptical trajectory in 
thin bilayer films
due to the demagnetizing field~\cite{optimum_spin_pump_ando},
\bege\label{eq_precession_around_y}
\!\hat{\vn{M}}(t)
\!=\!
\frac{1}{\eta(t)}\!\left[
\sin(\theta)\!\sin(\omega t)\epsilon,
\cos(\theta),
\sin(\theta)\!\cos(\omega t)
\right]^{\rm T},
\ee
where $\epsilon$ is 
the ratio of the major axis to the minor axis of the ellipse and
normalization of $\hat{\vn{M}}(t)$ is assured by
$\eta(t)\!=\!\sqrt{1\!+\![\epsilon^2\!-\!1]\!\sin^2(\omega t)\sin^2(\theta)}$.
The resulting induced current density is given by
\begin{gather}\label{eq_induced_current_precession_y}
\begin{aligned}
J_{x}^{\,\rm even}(t)&\!=\!
\frac{\omega\epsilon\sin^2\theta}{A\eta^2(t)}
\Bigl[
\!A_0\!+\!A_2
\frac{\eta^2(t)\!-\!\cos^2(\omega t)\sin^2\theta}{\eta^2(t)}
+\cdots
\Bigr]\\
&\!\!\!\!-\frac{\omega\epsilon\sin^2\theta\sin^2(\omega t)}{A\eta^4(t)}
\Bigl[
1+\sin^2\theta
(
\epsilon^2-1
)
\Bigr]
\Bigl[
B_2+\\
&\!\!\!\!+B_4
\frac{\eta^2(t)\!-\!\cos^2(\omega t)\sin^2\theta}{\eta^2(t)}
+\cdots
\Bigr],
\\
J_{x}^{\,\rm odd}(t)&=
\frac{\omega(1-\epsilon^2)}{2A\eta^3(t)}
\sin^2\theta\cos\theta\sin(2\omega t)
\Bigl[
C_0+\\
&+C_2
\frac{\eta^2(t)\!-\!\cos^2(\omega t)\sin^2\theta}{\eta^2(t)}
+\cdots\Bigr]\\
&-\frac{\omega\epsilon^2}{2A\eta^3(t)}\sin(2\omega t)\sin^2\theta\cos\theta
\Bigl[D_2+\\
&+D_4
\frac{\eta^2(t)\!-\!\cos^2(\omega t)\sin^2\theta}{\eta^2(t)}
+\cdots
\Bigr].
\end{aligned}\raisetag{7\baselineskip}
\end{gather}
For small angles $\theta$ the terms proportional to 
$\sin^2\theta$ dominate, while terms 
proportional to $\sin^4\theta$ and higher
are suppressed. Thus, we can approximate in the
small-cone limit
\begin{gather}
\begin{aligned}\label{eq_induced_currents_y}
J_{x}^{\,\rm even}(t)&=
\frac{\omega\epsilon}{A}\sin^2\theta
[
A_0+A_2+A_4+\cdots
]
\\
&-\frac{\omega\epsilon}{2A}\sin^2\theta
[
1-\cos(2\omega t)
]
[
B_2+B_4+\cdots
],
\\
J_{x}^{\,\rm odd}(t)&=
\frac{\omega}{2A}
\sin^2\theta\sin(2\omega t)(1-\epsilon^2)
[C_0+C_2+\cdots]\\
&-\frac{\omega}{2A}
\sin^2\theta
\sin(2\omega t)
\epsilon^2
\bigl[D_2+D_4
+\cdots\bigr].
\end{aligned}\raisetag{1.2\baselineskip}
\end{gather}
$J_{x}^{\,\rm even}$
is the sum of a dc component and an ac 
component with frequency $2\omega$, 
while 
$J_{x}^{\,\rm odd}$
consists of only an ac part with 
frequency $2\omega$. The ac components of the
even and odd part are phase shifted. Compared
to the induced current for precession around 
the $z$ axis, Eq.~\eqref{eq_induced_currents_z},
the amplitude is expected to be typically reduced by 
roughly a factor of $\sin\theta$ 
when the magnetization precesses around
the $y$ axis. The dc component of the
voltage $-R_{xx}J_{x}^{\,\rm even}L_y^{\phantom{y}}$, 
where 
$R_{xx}$ is the resistance, has been measured for
several bilayer systems and is usually 
interpreted as the voltage arising from
the conversion of pumped dc spin current via the 
ISHE~\cite{prl_mosendz_spin_pumping,
prb_mosendz_spin_pumping,prl_czeschka_spin_pumping}.

We turn now to the FMR-driven magnetization 
precession
around the $x$ axis. Again, the magnetization 
follows an elliptical
trajectory,
\bege
\hat{\vn{M}}(t)
=\frac{1}{\tilde{\eta}(t)}
\left[
\cos(\theta),
\sin(\theta)\cos(\omega t)\epsilon,
\sin(\theta)\sin(\omega t)
\right]^{\rm T},
\ee     
with $\epsilon$ the ratio of major axis 
to minor axis of the ellipse and
$\tilde{\eta}(t)\!=\!\sqrt{1\!+\![\epsilon^2\!-\!1]\!\cos^2(\omega t)\sin^2(\theta)}$.
In this case the current density induced 
by the precessing magnetization is
given by
\begin{gather}
\begin{aligned}
J_{x}^{\,\rm even}(t)&=
-\frac{\omega}{2A\tilde{\eta}^2(t)}
\sin(2\theta)
\cos(\omega t)
\Bigl\{  A_0+\\
&\!\!\!\!\!\!\!\!+
\frac{A_2}{\tilde{\eta}^2(t)}
[\cos^2\theta+\epsilon^2\sin^2\theta\cos^2(\omega t)]
+\cdots
\Bigr\}+\\
&\!\!\!\!\!\!\!\!+\frac{\omega}{2A\tilde{\eta}^4(t)}\sin(2\theta)\cos(\omega t)
\Bigl[
1+\sin^2\theta(\epsilon^2-1)
\Bigr]\times\\
&\!\!\!\!\!\!\!\!\times
\Bigl\{
B_2
+
\frac{B_4}{\tilde{\eta}^2(t)}
[\cos^2\theta+\epsilon^2\sin^2\theta\cos^2(\omega t)]
+\cdots
\Bigr\},\\
J_{x}^{\,\rm odd}(t)&=
-\frac{\omega\epsilon}{A\tilde{\eta}^3(t)}
\sin\theta\sin(\omega t)
\Bigl\{
C_0+\\
&\!\!\!\!\!\!\!\!\!\!\!\!+\frac{C_2}{\tilde{\eta}^2(t)}
[\cos^2\theta+\epsilon^2\sin^2\theta\cos^2(\omega t)]
+\cdots
\Bigr\}\\
&\!\!\!\!\!\!\!\!\!\!\!\!-\frac{\omega\epsilon}{A\tilde{\eta}^3(t)}
\sin\theta\cos^2\theta
\sin(\omega t)\Bigl\{
D_2+\\
&\!\!\!\!\!\!\!\!\!\!\!\!
+\frac{D_4}{\tilde{\eta}^2(t)}[\cos^2\theta+\epsilon^2\sin^2\theta\cos^2(\omega t)]
+\cdots
\Bigr\}.
\end{aligned}\raisetag{6\baselineskip}
\end{gather}
In the small-cone limit we obtain
\begin{gather}
\begin{aligned}\label{eq_induced_currents_x}
J_{x}^{\,\rm even}(t)&
\!=\!
\frac{\omega}{2A}\!
\sin(2\theta)\!
\cos(\omega t)\!
\Bigl[\!
B_2
\!+\!
B_4
\!+\!
\cdots
\!-\!
A_0
\!-\!
A_2
\!-\!
\cdots\!
\Bigr]\!=\\
&\!=\!
-{\torkance}_{yx}^{\rm even}(\hat{\vn{M}}=
\hat{\vn{e}}_{x}) \frac{\omega}{2A}
\sin(2\theta)
\cos(\omega t),\\
J_{x}^{\,\rm odd}(t)&
\!=\!
-\frac{\omega}{A}
\epsilon
\sin\theta\sin(\omega t)\!
\Bigl[\!
C_0
\!+\!
C_2
\!+
\cdots 
\!+\!
D_2
\!+\!
D_4
\!+
\cdots\!
\Bigr]\!=\\
&\!=\!
{\torkance}_{zx}^{\rm odd}(\hat{\vn{M}}=\hat{\vn{e}}_{x})
\frac{\omega}{A}
\epsilon
\sin\theta\sin(\omega t)
.
\end{aligned}\raisetag{1.2\baselineskip}
\end{gather}
Even if $A_2$, $B_2$, $C_2$ and $D_2$ are non-zero, i.e., even
in the presence of anisotropic SOT, the ISOT for magnetization
precession around the $x$ axis can thus be expressed 
in terms of the torkance
for magnetization along $x$. 
The even and odd contributions are again phase-shifted and the
dependence on the cone angle is $\propto\sin\theta$ in the
limit of small $\theta$ like in the case of magnetization precession
around the $z$ axis, promising a significantly larger 
ISOT signal~\cite{ac_voltage_spin_pumping_bauer} compared to the
case with magnetization precessing around the $y$ axis.

The main result of this subsection are the expressions 
for the ISOT currents given in
Eq.~\eqref{eq_induced_currents_z} (magnetization precession around $z$), 
Eq.~\eqref{eq_induced_currents_y} (magnetization precession around $y$)
and Eq.~\eqref{eq_induced_currents_x} (magnetization precession around $x$).
We stress that these expressions have been derived without any
assumptions on the underlying mechanism (such as SHE or interfacial
SOI) and are thus generally valid in bilayer systems with continuous
rotational symmetry around the $z$ axis.
In all three cases, the coefficients $C_0$, $C_2$, \dots and
$D_2$, $D_4$, \dots, which govern the odd torkance, give rise to an ac
current, but never to a dc current. Thus, complete characterization of
ISOT in experiments requires the measurement of the ac component.  

\subsection{Reciprocity between the even SOT and the even ISOT}
\label{sect_consequences_reciprocity}
In magnetic bilayer systems that involve a normal metal (NM) layer
with large SHE it is expected that an
important contribution
to the even SOT arises from SHE~\cite{stt_devices_giant_she_tungsten,
current_induced_switching_using_spin_torque_from_spin_hall_buhrman,
spin_torque_switching_TaCoFeB_Buhrman,perfect_alloys_spin_hall}. 
In particular when the NM layer is thin, the SHE 
in the NM layer will generally differ from the
SHE in a corresponding bulk system.
Even when the NM layer is thick, close to the 
interface with the magnet 
the electronic structure is
modified due to the hybridization of the electronic states 
of the NM with those of the ferromagnet (FM). This electronic
structure change is expected to entail a modification of the
SHE in the NM close to the interface. 
Furthermore, the proximity with the FM layer induces
magnetic moments in the NM at the interface
due to which the SHE is also modified~\cite{reduce_she_proximity}. 
Additionally, qualitatively new mechanisms for SHE are
added by the presence of the interface: When an
electric field is applied to the bilayer in the in-plane
direction, part of the
in-plane electric current is carried by interface states that
are evanescent waves along the stacking direction in the NM.  
That evanescent waves can also contribute to the SHE has
been discussed~\cite{she_evanescent_waves} in the context 
of tunnel junctions 
but is also true for evanescent waves in all-metallic bilayer systems.

Rather than using the term SHE only for the bulk contribution, 
we will in this 
work often denote by SHE the 
total spin current generated by an applied electric field,
including the interface modifications discussed above into the term SHE.
Recently, we have shown 
within \textit{ab initio} calculations that the even SOT in 
Co/Pt and Mn/W bilayers arises from the flux of spin current
from the NM into the FM layer~\cite{ibcsoit}.
Within our terminology this spin flux arises from the
SHE.

In the following we discuss a minimal model to describe the SHE-contribution
to the even SOT.
We consider a bilayer system composed of a semi-infinite ferromagnetic layer (FM)
on a semi-infinite normal metal (NM). The interface between FM
and NM is located at $z=0$.
We estimate the SOT arising from the SHE in NM,
when an electric field $E_{x}\hat{\vn{e}}_{x}$ is applied in
$x$ direction. 
Deep inside NM, i.e., for $z\ll 0$, the spin current
density flowing in $z$ direction is
\bege\label{eq_she_current_model}
Q^{\phantom{y}}_{y}
=
\sigma^{y}_{zx}E_{x}^{\phantom{y}}
=
\frac{\hbar}{2e}
\sigma_{xx}^{\phantom{y}}
E_{x}^{\phantom{y}}
\tan\gamma_{\rm SHE}^{\phantom{SHE}}
\ee
where $\sigma^{y}_{zx}$ is the SHE conductivity in NM,
$\tan\gamma_{\rm SHE}^{\phantom{SHE}}$ is the
SHE angle and $\sigma_{xx}^{\phantom{y}}$ is the normal
conductivity in NM. As discussed above 
the SHE is generally expected to be modified close to the
NM/FM interface. However, in order to obtain a minimal model we
neglect this expected position-dependence of SHE and assume
that the SHE can effectively be described by a single parameter $\sigma^{y}_{zx}$. 
We assume that
a fraction $\xi$ of $Q^{\phantom{y}}_{y}$ is transmitted through the
NM/FM interface 
and absorbed by FM, thereby causing a torque on its
magnetization, which we assume to point in $z$ direction.
This SHE-to-SOT efficiency $\xi$ can be thought of as the 
spin-current transmissivity~\cite{comparison_sot_spin_pumping_nifept}
or transparency~\cite{transparency_platinum_ferromagnet_interfaces} 
of the NM/FM interface. 
In metallic magnetic bilayer systems $\xi$ is typically of the order of 1:
In experiments on NiFe/Pt it was estimated 
to be $\xi\approx 0.4-0.6$~\cite{comparison_sot_spin_pumping_nifept}.
In \textit{ab initio} calculations of FePt/Pt $\xi\approx 0.6$ was 
found~\cite{fept_guillaume}.  
Denoting the $xy$ cross
sectional area of the unit cell by $A$, the torque per unit cell
is given by
${T}_{y}^{\rm even}
=
\xi A Q^{\phantom{y}}_{y}
=
{\torkance}_{yx}^{\rm even}
E^{\phantom{y}}_{x}$ with
\bege\label{eq_estimate_tyx_from_she}
{\torkance}_{yx}^{\rm even}(\hat{\vn{M}}=\hat{\vn{e}}_{z})
=
\xi A \sigma^{y}_{zx} =
\xi A \frac{\hbar}{2e}
\sigma_{xx}^{\phantom{y}}
\tan\gamma_{\rm SHE}^{\phantom{SHE}}.
\ee

Next, we consider the even ISOT arising from the
combined action of spin pumping and ISHE.
The spin current density pumped
adiabatically into NM
is determined
by~\cite{enhanced_gilbert_damping_ferromagnetic_films}
\bege\label{eq_spin_mixing_conduc}
\vn{Q}(z=0)=\frac{\hbar}{4\pi}
{\rm Re}
g^{\uparrow\downarrow}_{\phantom{xy}}
\hat{\vn{M}}\times \frac{d \hat{\vn{M}}}{d t},
\ee
where $g^{\uparrow\downarrow}_{\phantom{xy}}$ is the
(generally complex)
spin mixing conductance per cross-sectional area.
The imaginary part of $g^{\uparrow\downarrow}_{\phantom{xy}}$
is assumed to be negligible in
Eq.~\eqref{eq_spin_mixing_conduc}.
If spin transport in NM is diffusive, a spin accumulation
$s(z)$ forms in NM due to the spins pumped into NM.
The spin current in NM is proportional to the
gradient of the spin accumulation $s(z)$. Since the
spin accumulation decays exponentially
in NM, $s(z)=s(0)e^{z/\lambda_{\rm sd}}$,
where $\lambda_{\rm sd}$ is the spin diffusion length, also the
spin current decays exponentially in
NM, i.e., $\vn{Q}(z)=\vn{Q}(0)e^{z/\lambda_{\rm sd}}$~\cite{prl_mosendz_spin_pumping,ac_voltage_spin_pumping_bauer}.
In the case of magnetization precession around the
$y$ axis, Eq.~\eqref{eq_precession_around_y},
the dc spin current
flowing in NM in $z$ direction is therefore given by
\bege\label{eq_spin_current_profile_simplified}
Q^{\phantom{y}}_{y}(z)
=-\frac{\hbar\omega}{4\pi}
{\rm Re}
g^{\uparrow\downarrow}_{\phantom{xy}}
\sin^2(\theta)\,\epsilon \, e^{z/\lambda_{\rm sd}}.
\ee

Due to ISHE
this spin current is converted into an in-plane charge current
flowing in $x$ direction:
\bege\label{eq_charge_current_due_to_ishe_model}
\begin{aligned}
j ^{\rm even}_{x}(z)=&-
\frac{2e}{\hbar}Q^{\phantom{y}}_{y}(z)
\tan\gamma_{\rm ISHE}^{\phantom{SHE}}=\\
=&
\frac{e\omega}{2\pi}
{\rm Re}
g^{\uparrow\downarrow}_{\phantom{xy}}
\sin^2(\theta)\,\epsilon\,
e^{z/\lambda_{\rm sd}}
\tan\gamma_{\rm ISHE}^{\phantom{SHE}},
\end{aligned}
\ee
where $\tan\gamma_{\rm ISHE}^{\phantom{SHE}}$ is the ISHE-angle.
Thus, a single characteristic length,
the spin diffusion length $\lambda_{\rm sd}$, determines
the position dependence of $s(z)$, $Q_{y}(z)$ and $j ^{\rm even}_{x}(z)$
within this model:
\bege\label{eq_position_dependence_diffusive_model}
j_{x}^{\rm even}(z)\propto Q_{y}(z)\propto s(z) \propto e^{z/\lambda_{\rm sd}}.
\ee
Integration of the current
density Eq.~\eqref{eq_charge_current_due_to_ishe_model}
from $z=-\infty$ to $z=0$ yields the
current per length flowing in NM:
\bege
J^{\rm even}_{x}
=\frac{e\omega}{2\pi}
{\rm Re}
g^{\uparrow\downarrow}_{\phantom{xy}}
\sin^2(\theta)\,\epsilon\,
\lambda_{\rm sd}
\tan\gamma_{\rm ISHE}^{\phantom{SHE}}.
\ee
Using the small-cone
limit of Eq.~\eqref{eq_induced_current_precession_y}
and assuming $A_2=B_2=A_4=\dots=0$
we obtain the alternative
expression
\bege
J^{\rm even}_{x}
=
\frac{\omega}{A}\sin^2(\theta)\epsilon A_0.
\ee
Equating the two expressions
for $J^{\rm even}_{x}$
yields
\bege
A_{0}=A\frac{e}{2\pi}
{\rm Re}
g^{\uparrow\downarrow}_{\phantom{xy}}
\lambda_{\rm sd}
\tan\gamma_{\rm ISHE}^{\phantom{SHE}}.
\ee
Application of $t^{\rm even}_{yx}(\hat{\vn{M}}=\hat{\vn{e}}_{z})=A_{0}$ leads to
\bege\label{eq_lambda_ishe_connection}
\lambda_{\rm sd}=
\frac{
2\pi t^{\rm even}_{yx}(\hat{\vn{M}}=\hat{\vn{e}}_{z})
}
{
e A {\rm Re}
g^{\uparrow\downarrow}_{\phantom{xy}}
\tan\gamma_{\rm ISHE}^{\phantom{SHE}}
}.
\ee
Employing Eq.~\eqref{eq_estimate_tyx_from_she}
and
assuming $\tan\gamma_{\rm ISHE}^{\phantom{SHE}}=\tan\gamma_{\rm SHE}^{\phantom{SHE}}$
we can recast Eq.~\eqref{eq_lambda_ishe_connection} as
\bege\label{eq_intrinsic_lambda}
\lambda_{\rm sd}=\frac{\xi\hbar \pi\sigma_{xx}}{e^2 {\rm Re}
g^{\uparrow\downarrow}_{\phantom{xy}}}.
\ee
Eq.~\eqref{eq_lambda_ishe_connection} relates the SHE-to-SOT
efficiency $\xi$ with the parameters we use to model the ISOT current
and thereby expresses the reciprocity between SOT and ISOT. 

Even though this minimal model is derived for semiinfinite layers
it can be applied to bilayers of finite thickness when the layer
thickness is much larger than $\lambda_{\rm sd}$. 
When NM has the finite thickness $D$, i.e.,  $-D\le z \le 0$,
and when $D\gg \lambda_{\rm sd}$ is not satisfied, Eq.~\eqref{eq_spin_current_profile_simplified}
needs to be replaced by~\cite{prl_mosendz_spin_pumping}
\bege\label{eq_spin_current_profile_extended}
Q^{\phantom{y}}_{y}(z)
=-\frac{\hbar\omega}{4\pi}
{\rm Re}
g^{\uparrow\downarrow}_{\phantom{xy}}
\sin^2(\theta)\,\epsilon \, 
\frac{\sinh{\frac{z+D}{\lambda_{\rm sd}}}}
{\sinh{\frac{D}{\lambda_{\rm sd}}}},
\ee
in order to take into account that the spin current is reflected at
the boundary of NM at $z=-D$.

In Sec.~\ref{sec_even_sot} and~\ref{sec_even_isot} we will compare
ISOT current and spin current densities obtained 
from \textit{ab initio} 
calculations to the minimal model described
above. We will show that the minimal model provides a satisfactory
description of the \textit{ab initio} results. We will discuss that the main shortcoming of the minimal
model is the assumption that SHE and ISHE can be described by a
single position-independent parameter, whereby the modification of
SHE and ISHE close to the interface is neglected.
 
\subsection{Reciprocity between the odd SOT and the odd ISOT}
\label{sect_odd_boltzmann}
In Sec.~\ref{sec_relate_direct_and_inverse_sot}
we demonstrated the reciprocity between
ISOT and SOT on general grounds.
The odd SOT in the bilayer systems considered in this work 
arises dominantly from the
intraband contribution to Eq.~\eqref{eq_torkance_tensor}.
This intraband contribution can also be obtained from
Boltzmann transport theory within the
constant relaxation time approximation.
In this subsection we study the odd SOT and the odd ISOT
within Boltzmann transport theory and show that the
obtained expressions satisfy the reciprocity formulated previously
in Sec.~\ref{sec_relate_direct_and_inverse_sot}.

When an electric field $\vn{E}$ is applied the
occupation number $f^{\phantom{R}}_{\vn{k}n}$ of 
band $n$ at $k$-point $\vn{k}$
changes according to
\bege\label{eq_noneq_occupancy}
\delta f^{(1)}_{\vn{k}n}=-e\tau
\vn{v}^{\phantom{R}}_{\vn{k}n}
\cdot\vn{E}\,
\delta(\mathcal{E}^{\phantom{R}}_{\rm F}
-
\mathcal{E}^{\phantom{R}}_{\vn{k}n}),
\ee
where $\tau$ is the relaxation 
time, $\mathcal{E}^{\phantom{R}}_{\rm F}$ is the Fermi energy
and
$
\vn{v}^{\phantom{R}}_{\vn{k}n}=
\langle\psi^{\phantom{R}}_{\vn{k}n}
|
\vn{v}
|
\psi^{\phantom{R}}_{\vn{k}n}
\rangle
$
is the group velocity of band $n$ at $k$-point $\vn{k}$.
The change $\delta f^{(1)}_{\vn{k}n}$ of the occupancies
results in the contribution
\bege\label{eq_current_due_to_E}
\begin{aligned}
j^{(1)}_{\alpha}&=-\frac{e}{V\mathcal{N}}
\sum_{\vn{k}n}
v^{\phantom{R}}_{\vn{k}\alpha n}\delta f^{(1)}_{\vn{k}n}\\
&=\frac{e^2\tau}{V\mathcal{N}}
\sum_{\vn{k}n\beta}
v^{\phantom{R}}_{\vn{k}\alpha n}
v^{\phantom{R}}_{\vn{k}\beta n}
E_{\beta}^{\phantom{R}}\delta(\mathcal{E}^{\phantom{R}}_{\rm F}
-
\mathcal{E}^{\phantom{R}}_{\vn{k}n})
\end{aligned}
\ee
to the electric current density
and in the contribution
\bege\label{eq_torque_due_to_E}
\begin{aligned}
T^{(1)}_{\alpha}&=-
\sum_{\vn{k}n}
\mathcal{T}^{\phantom{R}}_{\vn{k}\alpha n}
\delta f^{(1)}_{\vn{k}n}\\
&=\frac{e\tau}{\mathcal{N}}\sum_{\vn{k}n\beta}
\mathcal{T}^{\phantom{R}}_{\vn{k}\alpha n}
v^{\phantom{R}}_{\vn{k}\beta n}
E_{\beta}^{\phantom{R}}\delta(\mathcal{E}^{\phantom{R}}_{\rm F}
-
\mathcal{E}^{\phantom{R}}_{\vn{k}n}),
\end{aligned}
\ee
to the torque,
where $v^{\phantom{R}}_{\vn{k}\alpha n}$ 
and $\mathcal{T}^{\phantom{R}}_{\vn{k}\alpha n}$ 
are the $\alpha$-th Cartesian components of the
group velocity $\vn{v}^{\phantom{R}}_{\vn{k}n}$ and of the 
torque $\vn{\mathcal{T}}^{\phantom{R}}_{\vn{k}n}=
\langle\psi^{\phantom{R}}_{\vn{k}n}
|
\vn{\mathcal{T}}
|
\psi^{\phantom{R}}_{\vn{k}n}
\rangle$, respectively. 

When the system is perturbed not by an electric field but
by the time-dependence of the magnetization 
direction $\hat{\vn{M}}(t)$ the change of the
occupancies is given by
\bege\label{eq_noneq_occupation_mdot}
\delta f^{(2)}_{\vn{k}n}=\tau
\delta(\mathcal{E}^{\phantom{R}}_{\rm F}
-
\mathcal{E}^{\phantom{R}}_{\vn{k}n})
\vn{\mathcal{T}}^{\phantom{R}}_{\vn{k}n}
\cdot
\left[
\hat{\vn{M}}(t)
\times
\frac{d \hat{\vn{M}}(t)}{d\,t}
\right]
\ee
instead of Eq.~\eqref{eq_noneq_occupancy}.
Eq.~\eqref{eq_noneq_occupation_mdot} follows from
\bege\label{eq_noneq_occupation_mdot2}
\begin{aligned}
&\frac{\delta f^{(2)}_{\vn{k}n}}{\tau}
=-\frac{\partial f^{\phantom{R}}_{\vn{k}n}}{\partial \hat{\vn{M}}}
\cdot\frac{d \hat{\vn{M}}}{d\,t}
=-\frac{\partial f^{\phantom{R}}_{\vn{k}n}}
{\partial \mathcal{E}^{\phantom{R}}_{\vn{k}n}}
\frac{\partial\mathcal{E}^{\phantom{R}}_{\vn{k}n}}
{\partial \hat{\vn{M}}}
\cdot\frac{d \hat{\vn{M}}}{d\,t}=\\
&=
\delta(\mathcal{E}^{\phantom{R}}_{\rm F}
\!-\!
\mathcal{E}^{\phantom{R}}_{\vn{k}n})
\left[
\hat{\vn{M}}(t)
\!\times\!
\frac{\partial\mathcal{E}^{\phantom{R}}_{\vn{k}n}}
{\partial \hat{\vn{M}}}
\right]
\!\cdot\!
\left[
\hat{\vn{M}}(t)
\!\times\!
\frac{d \hat{\vn{M}}(t)}{d\, t}
\right]\\
&=\delta(\mathcal{E}^{\phantom{R}}_{\rm F}
\!-\!
\mathcal{E}^{\phantom{R}}_{\vn{k}n})
\vn{\mathcal{T}}^{\phantom{R}}_{\vn{k}n}
\!\cdot\!
\left[
\hat{\vn{M}}(t)
\!\times\!
\frac{d \hat{\vn{M}}(t)}{d\, t}
\right],\\
\end{aligned}
\ee
where we set the temperature in the Fermi-Dirac 
distribution function
to zero such 
that $f^{\phantom{R}}_{\vn{k}n}=\theta(\mathcal{E}^{\phantom{R}}_{\rm F}
\!-\!
\mathcal{E}^{\phantom{R}}_{\vn{k}n})$ 
and $\partial f^{\phantom{R}}_{\vn{k}n}/\partial \mathcal{E}^{\phantom{R}}_{\vn{k}n} =-\delta(\mathcal{E}^{\phantom{R}}_{\rm F}
\!-\!
\mathcal{E}^{\phantom{R}}_{\vn{k}n})$. 
Additionally,
we made use of
\bege
\left[
\hat{\vn{M}}(t)
\!\times\!
\frac{\partial\mathcal{E}^{\phantom{R}}_{\vn{k}n}}
{\partial \hat{\vn{M}}}
\right]
\!=\!
\hat{\vn{M}}(t)
\!\times\!
\left\langle
\!
\psi^{\phantom{R}}_{\vn{k}n}
\!
\left|
\frac{\partial H_{\hat{\vn{M}}}}
{\partial \hat{\vn{M}}}
\right|
\!
\psi^{\phantom{R}}_{\vn{k}n}
\!
\right\rangle
\!=\!
\vn{\mathcal{T}}^{\phantom{R}}_{\vn{k}n}.
\ee
Eq.~\eqref{eq_noneq_occupation_mdot} 
and Eq.~\eqref{eq_noneq_occupation_mdot2} hold under the
condition that the frequency $\omega$ of the precession 
of magnetization is small compared to the relaxation rate
$\tau^{-1}$, i.e., $\omega\ll \tau^{-1}$. If the condition
$\omega\ll \tau^{-1}$ is violated one needs to solve the
Boltzmann equation assuming an explicit time-dependence 
of the distribution function. The expressions valid in
that case are obtained by replacing $\tau$ in 
Eq.~\eqref{eq_noneq_occupation_mdot} 
and Eq.~\eqref{eq_noneq_occupation_mdot2} as follows:
\bege
\tau\rightarrow\frac{\tau}{1-i\omega\tau}.
\ee
For magnetic bilayers such as Co/Pt we estimate that
$10{\rm THz}<1/(2\pi\tau)$, which is much larger than 
ferromagnetic resonance frequencies in the GHz range. 
Therefore, we will always assume $\omega\ll \tau^{-1}$ in the following.

The current density induced due to the time-dependence 
of magnetization
can be obtained from 
the change of occupancies $\delta f^{(2)}_{\vn{k}n}$ given 
in Eq.~\eqref{eq_noneq_occupation_mdot}:
\begin{gather}\label{eq_induced_current_intraband}
\begin{aligned}
j^{(2)}_{\alpha}&=-\frac{e}{V\mathcal{N}}
\sum_{\vn{k}n}
v^{\phantom{R}}_{\vn{k}\alpha n}\delta f^{(2)}_{\vn{k}n}\\
&=-\frac{e\tau}{V\mathcal{N}}
\sum_{\vn{k}n}\!
v^{\phantom{R}}_{\vn{k}\alpha n}
\delta(\mathcal{E}^{\phantom{R}}_{\rm F}
\!-\!
\mathcal{E}^{\phantom{R}}_{\vn{k}n})
\vn{\mathcal{T}}^{\phantom{R}}_{\vn{k}n}
\!\cdot\!
\left[\!
\hat{\vn{M}}(t)
\!\times\!
\frac{d \hat{\vn{M}}(t)}{d\, t}
\!\right].
\end{aligned}\raisetag{4\baselineskip}
\end{gather}
Similarly, the torque which damps the magnetization dynamics
is given by
\begin{gather}\label{eq_damping_intraband}
\begin{aligned}
&T^{(2)}_{\alpha}=-\frac{1}{\mathcal{N}}
\sum_{\vn{k}n}
\mathcal{T}^{\phantom{R}}_{\vn{k}\alpha n}
\delta f^{(2)}_{\vn{k}n}=\\
&=-\frac{\tau}{\mathcal{N}}
\sum_{\vn{k}n}
\mathcal{T}^{\phantom{R}}_{\vn{k}\alpha n}
\delta(\mathcal{E}^{\phantom{R}}_{\rm F}
-
\mathcal{E}^{\phantom{R}}_{\vn{k}n})
\vn{\mathcal{T}}^{\phantom{R}}_{\vn{k}n}
\cdot
\left[
\hat{\vn{M}}(t)
\!\times\!
\frac{d \hat{\vn{M}}(t)}{d\, t}
\right].
\end{aligned}\raisetag{4\baselineskip}
\end{gather}

We can combine Eq.~\eqref{eq_current_due_to_E}, 
Eq.~\eqref{eq_torque_due_to_E}, 
Eq.~\eqref{eq_induced_current_intraband} and
Eq.~\eqref{eq_damping_intraband} in the form of
Eq.~\eqref{eq_a_matrix_linear_response} as follows:
\bege\label{eq_a_intra_linear_response}
\begin{aligned}
\begin{pmatrix}
\tilde{\vn{j}}\\
\tilde{\vn{T}}/V
\end{pmatrix}=
&
\begin{pmatrix}
\tilde{\vht{\sigma}} &-\tilde{\vn{\torkance}}^{\rm T}/V\\
\tilde{\vn{\torkance}}/V & -\tilde{\vht{\Lambda}}
\end{pmatrix}
\begin{pmatrix}
\vn{E}\\
\hat{\vn{M}}\times \frac{d \hat{\vn{M}}}{dt}
\end{pmatrix}\\
\end{aligned},
\ee
where we defined $\tilde{\vn{j}}=\vn{j}^{(1)}+\vn{j}^{(2)}$ 
and $\tilde{\vn{T}}=\vn{T}^{(1)}+\vn{T}^{(2)}$. We use the
tilde to recall that according to Eq.~\eqref{eq_current_due_to_E}, 
Eq.~\eqref{eq_torque_due_to_E}, 
Eq.~\eqref{eq_induced_current_intraband} and
Eq.~\eqref{eq_damping_intraband} only intraband terms are
considered in $\tilde{\vn{j}}$ and $\tilde{\vn{T}}$, while the
complete expression for current density and torque contains
additional interband terms. The linear response coefficients
$\tilde{\vht{\sigma}}$, $\tilde{\vn{\torkance}}$ 
and $\tilde{\vht{\Lambda}}$ are given by
\bege\label{eq_list_intraband_coefficients}
\begin{aligned}
\tilde{\sigma}^{\phantom{R}}_{\alpha\beta}&=\frac{e^2\tau}{V\mathcal{N}}
\sum_{\vn{k}n}
v^{\phantom{R}}_{\vn{k}\alpha n}
v^{\phantom{R}}_{\vn{k}\beta n}
\delta(\mathcal{E}^{\phantom{R}}_{\rm F}
-
\mathcal{E}^{\phantom{R}}_{\vn{k}n}),\\
\tilde{\torkance}^{\phantom{R}}_{\alpha\beta}&=
\frac{e\tau}{\mathcal{N}}
\sum_{\vn{k}n}
\mathcal{T}^{\phantom{R}}_{\vn{k}\alpha n}
v^{\phantom{R}}_{\vn{k}\beta n}
\delta(\mathcal{E}^{\phantom{R}}_{\rm F}
-
\mathcal{E}^{\phantom{R}}_{\vn{k}n}),\\
\tilde{\Lambda}^{\phantom{R}}_{\alpha\beta}&=\frac{\tau}{V\mathcal{N}}
\sum_{\vn{k}n}
\mathcal{T}^{\phantom{R}}_{\vn{k}\alpha n}
\mathcal{T}^{\phantom{R}}_{\vn{k}\beta n}
\delta(\mathcal{E}^{\phantom{R}}_{\rm F}
-
\mathcal{E}^{\phantom{R}}_{\vn{k}n}).
\end{aligned}
\ee
$\tilde{\vht{\sigma}}$ and $\tilde{\vht{\Lambda}}$
are even with respect to reversal of 
magnetization direction $\hat{\vn{M}}$,
while $\tilde{\vn{\torkance}}$ is odd. 
Eq.~\eqref{eq_a_intra_linear_response} clearly shows that the 
tensor $\tilde{\vn{\torkance}}$ governs both the
odd SOT and the odd ISOT. 
The Gilbert damping $\tilde{\vn{\alpha}}$ is 
related to $\tilde{\vn{\Lambda}}$ 
by $\tilde{\vn{\alpha}}=|\gamma|\tilde{\vn{\Lambda}}/(\mu_0 M)$ 
(see Eq.~\eqref{eq_gilb_damp}), i.e.,
\bege
\tilde{\alpha}^{\phantom{R}}_{\alpha\beta}=
\frac{|\gamma|\tau}{\mu_0 VM\mathcal{N}}
\sum_{\vn{k}n}
\mathcal{T}^{\phantom{R}}_{\vn{k}\alpha n}
\mathcal{T}^{\phantom{R}}_{\vn{k}\beta n}
\delta(\mathcal{E}^{\phantom{R}}_{\rm F}
-
\mathcal{E}^{\phantom{R}}_{\vn{k}n}),
\ee
which agrees with the intraband term in the torque-correlation formula
of the Gilbert 
damping~\cite{damping_gilmore_stiles,
damping_kunes_kambersky,damping_garate_macdonald}.
Thus, Eq.~\eqref{eq_noneq_occupation_mdot} leads to a coherent
description of the intraband contributions to both the Gilbert damping
and the odd ISOT. Moreover, the expression obtained for the odd ISOT 
is reciprocal to the odd direct SOT.

\section{First principles calculations}
\label{sec_first_principles}
\subsection{Computational method}
In the following we will discuss SOTs and ISOTs for
a bilayer composed of
3 layers of hcp Co on 20 layers of fcc Pt(111),
denoted in the following as Co(3)/Pt(20).
We label the atomic layers of the Pt
layer by Pt1 through Pt20, where
Pt20 is at the Co/Pt interface. Likewise, we label
the atomic layers of the Co layer by
Co1 through Co3, where Co1 is at
the Co/Pt interface.
We introduce a cartesian coordinate system such that
the $z$ axis is perpendicular to the atomic layers,
i.e., along the out-of-plane direction, and Pt20 has a smaller
$z$ coordinate than Co1. The magnetization direction
is set to $\hat{\vn{M}}=\hat{\vn{e}}_{z}$ in the calculation.
In order to perform the linear-response calculations of the torkance
computationally
efficiently, the Wannier interpolation technique 
is 
employed~\cite{WannierPaper,wannier90,rmp_wannier90}. For this purpose
we express the electronic structure in terms of maximally localized
Wannier functions (MLWFs), using 18 MLWFs per atom.
Details of the electronic structure calculation of Co(3)/Pt(20) are given
in Ref.~\cite{ibcsoit}. 

Within the independent particle approximation 
the torkance $\vn{\torkance}$
defined in Eq.~\eqref{eq_torkance_tensor} can 
be expressed as sum of 
three terms, ${\torkance}^{\phantom{I}}_{\alpha\beta}=
{\torkance}^{\rm I(a)}_{\alpha\beta}
+
{\torkance}^{\rm I(b)}_{\alpha\beta}
+
{\torkance}^{\rm II}_{\alpha\beta}$, 
where~\cite{ibcsoit,mothedmisot} 
\begin{gather}\label{eq_kubo_linear_response_torkance}
\begin{aligned}
{\torkance}^{\rm I(a)\phantom{I}}_{\alpha\beta}\!\!\!\!&=
\phantom{-}\frac{e}{\mathcal{N}h}\sum_{\vn{k}}
\,{\rm Tr}
\left\langle
\mathcal{T}_{\alpha}
G^{\rm R}_{\vn{k}}(\mathcal{E}_{\rm F})
v_{\beta}
G^{\rm A}_{\vn{k}}(\mathcal{E}_{\rm F})
\right\rangle
\\
{\torkance}^{\rm I(b)\phantom{I}}_{\alpha\beta}\!\!\!\!&=
-\frac{e}{\mathcal{N}h}\sum_{\vn{k}}
\,{\rm Re}
\,{\rm Tr}
\left\langle
\mathcal{T}_{\alpha}
G^{\rm R}_{\vn{k}}(\mathcal{E}_{\rm F})
v_{\beta}
G^{\rm R}_{\vn{k}}(\mathcal{E}_{\rm F})
\right\rangle
\\
{\torkance}^{\rm II\phantom{(a)}}_{\alpha\beta}\!\!\!\!&=
\phantom{-}\frac{e}{\mathcal{N}h}\sum_{\vn{k}}\int_{-\infty}^{\mathcal{E}_{\rm F}}
d\mathcal{E}
\,{\rm Re}
\,{\rm Tr}
\left\langle
\mathcal{T}_{\alpha}G^{\rm R}_{\vn{k}}(\mathcal{E})v_{\beta}
\frac{dG^{\rm R}_{\vn{k}}(\mathcal{E})}{d\mathcal{E}}\right.\\
 &\quad\quad\quad\quad\quad\quad\quad\quad\,-\left.
\mathcal{T}_{\alpha}\frac{dG^{\rm R}_{\vn{k}}(\mathcal{E})}{d\mathcal{E}}v_{\beta}G^{\rm R}_{\vn{k}}(\mathcal{E})
\right\rangle,
\end{aligned}\raisetag{5.6\baselineskip}
\end{gather}
with $G^{\rm R}_{\vn{k}}(\mathcal{E})$
the retarded Green function at $k$ 
point $\vn{k}$ and energy $\mathcal{E}$,
$G^{\rm A}_{\vn{k}} (\mathcal{E})$ the
advanced one, $\mathcal{N}$ the number of $k$ points and
$\mathcal{E}_{\rm F}$ the Fermi energy.
We model the effect of disorder by a phenomenological
band broadening $\Gamma$ in the Green functions, i.e.,
$G^{\rm R}_{\vn{k}}(\mathcal{E})=\hbar[\mathcal{E}-H_{\vn{k}}+i\Gamma]^{-1}$. 

We discuss the direct SOT in terms of the torkance, which we compute
according to Eq.~\eqref{eq_kubo_linear_response_torkance}.
In order to obtain atom-resolved torkances,
we replace the torque operator 
in Eq.~\eqref{eq_kubo_linear_response_torkance}
by an atom-resolved torque operator (see Ref.~\cite{ibcsoit} for details).
We calculate 
the induced ISOT current
in the Co(3)/Pt(20) bilayer using 
Eq.~\eqref{eq_even_odd_currents} and
the torkance obtained 
from Eq.~\eqref{eq_kubo_linear_response_torkance}.
However, it is desirable to determine also the spatial profile
of the ISOT current along the $z$ direction.
For this purpose we define
the layer-resolved velocity operator
\bege\label{eq_modified_velocity_operator}
v_{\vn{k}\alpha nm}(L)=
v_{\vn{k}\alpha nm}
\theta_n(L)\theta_m(L),
\ee 
where $\theta_m(L)=1$ if MLWF orbital $m$ belongs to 
layer $L$ and zero otherwise.
Here, each MLWF is attributed to the one atomic layer 
in which the center of the MLWF is located 
and
\bege
v_{\vn{k}\alpha nm}=\frac{1}{\hbar}\sum_{\vn{R}}
e^{i\vn{k}\cdot\vn{R}}i{\rm R}_{\alpha}
\langle
W_{n\vn{0}}
|
H
|
W_{m\vn{R}}
\rangle
\ee
is the $\alpha$-th cartesian component of the 
velocity operator at $k$ point $\vn{k}$ expressed in the
basis of Wannier functions.
Replacing $v_{\alpha}$ in 
Eq.~\eqref{eq_kubo_linear_response_torkance} by
$v_{\alpha}(L)$ allows us to compute the ISOT current
within the atomic layer $L$. 

The direct SOT is a response to the applied electric field $\vn{E}$,
which exerts the mechanical force $-e\vn{E}$ on the electrons.
By artificially switching off the force $-e\vn{E}$ for some atomic 
layers, we investigate which atomic layers participate in 
generating the SOT.
Noting that the
mechanical force is represented
in Eq.~\eqref{eq_kubo_linear_response_torkance}
by the velocity
operator, we replace $v_{\alpha}$ in 
Eq.~\eqref{eq_kubo_linear_response_torkance} by
$v_{\alpha}(L)$ in order to study the SOT generated when the force
$-e\vn{E}$ acts only on the electrons in the atomic layer $L$.
Thus, the replacement of $v_{\alpha}$ by $v_{\alpha}(L)$ in
Eq.~\eqref{eq_kubo_linear_response_torkance} provides us not only
with the information on how the ISOT current is distributed in Co(3)/Pt(20)
along the stacking direction, but 
additionally it also provides us with the information in which atomic
layers the action of the force $-e\vn{E}$ is essential for the direct SOT. 
This results from the reciprocity between
ISOT and SOT, which implies that the atomic layers that 
carry the ISOT current 
agree to the atomic layers that participate in 
generating the SOT.  
In order to describe the situation where the
mechanical force is switched off 
for the atomic layers Pt1 through $L-1$ we use the sum of
Eq.~\eqref{eq_modified_velocity_operator} for the layers $L,L+1,\dots$,
i.e., we use the modified velocity operator 
\bege\label{eq_mech_force_off_region}
\bar{v}_{\vn{k}\alpha n m}(L)=v_{\vn{k}\alpha n m}
\sum_{L_1\ge L}
\sum_{L_2\ge L}\theta_{n}(L_1)\theta_{m}(L_2)
\ee
in Eq.~\eqref{eq_kubo_linear_response_torkance}.
Here, the
functions $\theta_{m}(L)$ are defined
like above,
below Eq.~\eqref{eq_modified_velocity_operator}.

As discussed in Sec.~\ref{sect_consequences_reciprocity},
the spin current flowing in $z$ direction mediates an important 
contribution to the
even ISOT in bilayer systems. Thus, it is 
desirable to determine its spatial profile along
the $z$ direction. 
For this purpose, we define the layer-resolved 
spin current density 
operator $\mathcal {Q}_{s}(L)$
for spin currents flowing in $z$ direction
by
\bege\label{eq_integrated_current_density}
\langle\psi_{\vn{k}n}|\mathcal{Q}_{s}(L)|\psi_{\vn{k}m}\rangle\!=\!
\frac{1}{A}\!\int_{S_{L}} \!\!d\vn{S}\!\cdot\!
\langle\psi_{\vn{k}n}|
\vht{\mathcal{Q}}_{s}(\vn{r})
|\psi_{\vn{k}m}\rangle
,
\ee 
where the integration is 
over the boundary $S_{L}$ between layers $L-1$ and $L$,
$A$ is the $xy$ cross sectional area of the 
unit cell, and $\vht{\mathcal{Q}}_{s}(\vn{r})$ is the
spin current density operator at point $\vn{r}$.
SOI is only strong close to the atomic nuclei, because it is
proportional to the electrostatic potential gradient.
Since the boundary $S_{L}$ is chosen to lie in the interstitial region,
where SOI is negligible, the nonrelativistic
spin current density operator can be used:
\bege\label{eq_spin_current_density}
\vht{\mathcal{Q}}_{s}(\vn{r})
=
\frac{\hbar}{2}\frac{\hbar}{2im}\Bigl[
\delta(\vn{r}-\hat{\vn{r}})\overset{\Rightarrow}{\nabla}
-\overset{\Leftarrow}{\nabla}\delta(\vn{r}-\hat{\vn{r}})
\Bigr]\sigma_{s}.
\ee
By replacing in Eq.~\eqref{eq_induced_current} 
the current density operator $-ev_{\alpha}/V$ by 
$\mathcal{Q}_{s}(L)$, we can determine the 
spin current profile along the stacking direction of the
Co(3)/Pt(20) bilayer:
\bege\label{eq_induced_spin_current}
{Q}_{s}(L,t)\!=\!
\frac{1}{A}
\!\sum_{\beta}
\!w_{s\beta}(L,\hat{\vn{M}}(t))
 \!\!\left[
 \!\hat{\vn{M}}(t)
 \!\times\!
 \frac{d\hat{\vn{M}}(t)}{dt}
 \!\right]_{\beta},
\ee
where we defined
\bege\label{eq_layer_resolved_spin_current}
w_{s\beta}(L,\hat{\vn{M}})=-
A\lim_{\omega\to 0}
\frac{
{\rm Im}
G_{
\mathcal{Q}_{s}(L),
\mathcal{T}_{\beta}
}
^{\rm R}
(
\hbar\omega,\hat{\vn{M}}
)
}
{\hbar\omega},
\ee
with the Fourier transform of the retarded spin-current 
torque correlation function
\bege
\!\!G_{\mathcal{Q}_{s}(L),\mathcal{T}_{\beta}}^{\rm R}
(\hbar\omega,\hat{\vn{M}})\!=\!-i\!\!\int\limits_{0}^{\infty}\!\!dt 
e^{i\omega t}\!
\left\langle
[
\mathcal{Q}_{s}(L),\!\mathcal{T}_{\beta}(-t)
]_{-}
\right\rangle.
\ee
Within the independent particle 
approximation Eq.~\eqref{eq_layer_resolved_spin_current}
becomes 
$
w^{\phantom{I}}_{s\beta}(L)
\!=
w^{\rm I(a)}_{s\beta}(L)
\!+
w^{\rm I(b)}_{s\beta}(L)
\!+
w^{\rm II}_{s\beta}(L)
$, with
\begin{gather}\label{eq_kubo_linear_response_pumped_spin_current}
\begin{aligned}
w^{\rm I(a)\phantom{I}}_{s\beta}\!\!\!(L)\!
&=\frac{eA}{\mathcal{N}h}\!\sum_{\vn{k}}
{\rm Tr}
\left\langle
\mathcal{Q}_{s}(L)
G^{\rm R}_{\vn{k}}(\mathcal{E}_{\rm F})
\mathcal{T}_{\beta}
G^{\rm A}_{\vn{k}}(\mathcal{E}_{\rm F})
\right\rangle
\\
w^{\rm I(b)\phantom{I}}_{s\beta}\!\!\!(L)\!
&=-\frac{eA}{\mathcal{N}h}\!\sum_{\vn{k}}
{\rm Re}
\,{\rm Tr}
\left\langle
\mathcal{Q}_{s}(L)
G^{\rm R}_{\vn{k}}(\mathcal{E}_{\rm F})
\mathcal{T}_{\beta}
G^{\rm R}_{\vn{k}}(\mathcal{E}_{\rm F})
\right\rangle
\\
w^{\rm II\phantom{(a)}}_{s\beta}\!\!\!(L)\!
&=
\frac{eA}{\mathcal{N}h}
\!\sum_{\vn{k}}
\!\int_{-\infty}^{\mathcal{E}_{\rm F}}
\!d\mathcal{E}
\,{\rm Re}
\,{\rm Tr}
\left\langle
\!\mathcal{Q}_{s}(L)G^{\rm R}_{\vn{k}}(\mathcal{E})\mathcal{T}_{\beta}
\frac{dG^{\rm R}_{\vn{k}}(\mathcal{E})}{d\mathcal{E}}\right.\\
 &\,-\left.
\mathcal{Q}_{s}(L)\frac{dG^{\rm R}_{\vn{k}}(\mathcal{E})}{d\mathcal{E}}
\mathcal{T}_{\beta}
G^{\rm R}_{\vn{k}}(\mathcal{E})
\right\rangle,
\end{aligned}\raisetag{1.3\baselineskip}
\end{gather}
where we suppressed the $\hat{\vn{M}}$ dependence 
for notational
convenience. Comparison of Eq.~\eqref{eq_spin_mixing_conduc} and
Eq.~\eqref{eq_induced_spin_current}
yields the following expression for the spin mixing conductance:
\bege\label{eq_mix_fp}
{\rm Re}g^{\uparrow\downarrow}_{\phantom{xy}}
=\frac{4\pi}{\hbar A}
w_{yy}(L={\rm Co1}),
\ee
where $w_{yy}(L={\rm Co1})$ is proportional
to spin current flowing between the layers Pt20 and Co1.
In Co/Pt bilayers ${\torkance}^{\rm even}_{yx}$ arises almost entirely
from the spin-flux into the Co-layer~\cite{ibcsoit}.
The extraction
of ${\rm Re}g^{\uparrow\downarrow}$ from $w_{yy}$ is therefore
meaningful in this case despite the presence of
SOI in the calculation.

Similarly, as discussed in Sec.~\ref{sect_consequences_reciprocity},
SHE provides an important contribution to the even SOT in bilayer systems.
The spin currents of the direct SHE are generated
by the applied electric field rather than by
spin pumping. 
In order to investigate the layer-resolved spin current profile 
of these spin currents in Co(3)/Pt(20),
we define the coefficients
\bege\label{eq_layer_resolved_spin_current_sot}
{q}_{s\beta}(L,\hat{\vn{M}})=Ae
\lim_{\omega\to 0}
\frac{
{\rm Im}
G_{
{\mathcal{Q}}_{s}(L),
v_{\beta}
}
^{\rm R}
(
\hbar\omega,\hat{\vn{M}}
)
}
{\hbar\omega}.
\ee
For example, ${q}_{yx}(L)$ quantifies 
the linear response of 
spin currents flowing in $z$ direction with spin
pointing in $y$ direction to the electric 
field in $x$ direction. 
Within the independent particle approximation
${q}_{s\beta}(L)$ is expressed similarly to the
torkance (Eq.~\eqref{eq_kubo_linear_response_torkance}):
Only $\mathcal{T}_{\alpha}$ has to be replaced by
$-A{Q}_{s}(L)$ in the expressions.

For a given atomic layer,
the difference between spin-current flowing in
and spin-current flowing out 
is the spin-flux into that atomic layer.
In Co/Pt bilayer systems, the even SOT arises 
dominantly from the spin-flux into the Co layer~\cite{ibcsoit}.
The
linear-response coefficient of
spin flux into layer $L$ is given by
\bege\label{eq_spin_current_flux_even}
\Delta q_{yx}(L)=q_{yx}(L)
-q_{yx}(L+1),
\ee
where according
to Eq.~\eqref{eq_integrated_current_density}
and Eq.~\eqref{eq_layer_resolved_spin_current_sot}
$q_{yx}(L)$ describes spin current flowing
between layers $L-1$ and $L$ towards layer $L$
and $-q_{yx}(L+1)$ describes spin current
flowing between layers $L$ and $L+1$ towards layer $L$.

\subsection{Even SOT}
\label{sec_even_sot}
We first discuss the even torkance ${\torkance}_{yx,\rm\,25meV}^{\rm even}$
determined from Eq.~\eqref{eq_kubo_linear_response_torkance}.
At $\Gamma$~=~25~meV we 
obtain ${\torkance}_{yx,\rm\,25meV}^{\rm even}=0.68\,ea_{0}$ 
per unit cell, 
where $ea_{0}$ is the
atomic unit of torkance, which 
amounts to $ea_{0}=8.478\cdot 10^{-30}$~Cm.
A slightly smaller value 
of ${\torkance}_{yx,\rm\,100meV}^{\rm even}=0.53\,ea_{0}$ is 
calculated at $\Gamma$~=~100~meV.
Dividing these torkances by the magnetic 
moment per unit cell 
of $\mu=5.78 \mu_{\rm B}$ we compute the 
effective fields per applied
electric field 
of ${\torkance}_{yx,\rm\,25meV}^{\rm even}/\mu=0.011\,{\rm mTcm/V}$
and ${\torkance}_{yx,\rm\,100meV}^{\rm even}/\mu=0.0084\,{\rm mTcm/V}$.

\begin{figure}
\includegraphics*[width=8.5cm]{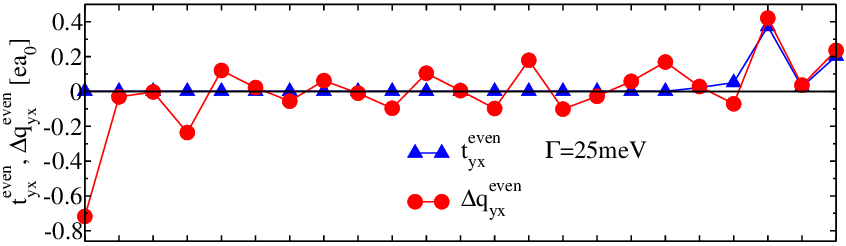}
\includegraphics*[width=8.7cm]{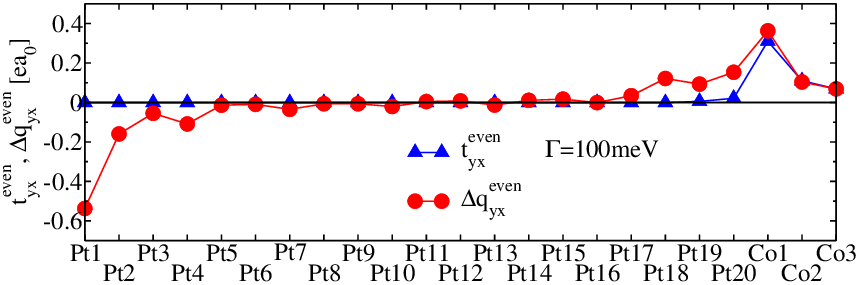}
\caption{\label{fig_layer_resolved_even_torkance}
Triangles: Layer-resolved even
torkance ${\torkance}_{yx}^{\rm even}$
for broadenings of
$\Gamma=25$~meV (upper graph)
and  $\Gamma=100$~meV (lower graph).
Circles: Linear response coefficient of the
layer-resolved spin flux $\Delta q_{yx}^{\rm even}$,
Eq.~\eqref{eq_spin_current_flux_even}.
Solid lines serve as guide to the eye.
}
\end{figure}

In Fig.~\eqref{fig_layer_resolved_even_torkance}
we show the
layer-resolved even torkance, i.e., the
linear-response coefficient of the torque
acting on the magnetization of a given layer, and the
linear-response coefficient of
spin flux  
into layer $L$ (Eq.~\eqref{eq_spin_current_flux_even}).
For the Co layers, layer-resolved torkances and
spin fluxes coincide approximately.
Thus, the even torkance in Co(3)/Pt(20) arises dominantly
from the spin current flowing into the Co layer,
consistent with the discussion in 
Sec.~\ref{sect_consequences_reciprocity} and with previous
work on Co/Pt bilayer systems~\cite{ibcsoit}.

\begin{figure}
\includegraphics*[width=8.7cm]{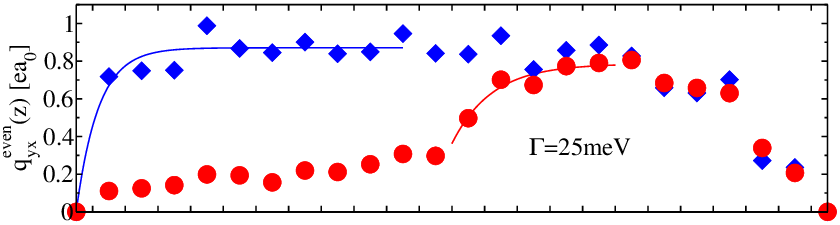}
\includegraphics*[width=8.7cm]{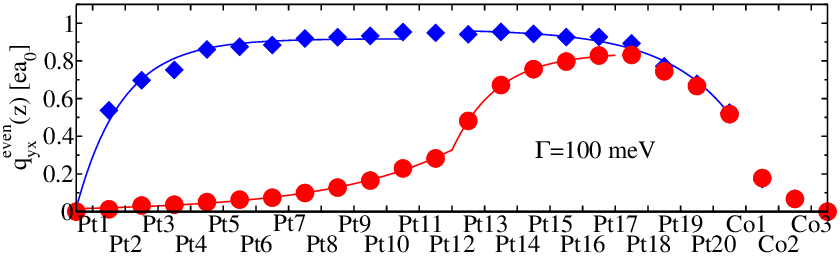}
\caption{\label{fig_even_torque_mechanical_force_on_off}
Diamonds: Linear response
coefficients ${q}_{yx}^{\rm even}(z)$
of
the layer-resolved
spin current for $\Gamma=25$~meV (upper graph)
and  $\Gamma=100$~meV (lower graph).
Circles: Linear response coefficients
${q}_{yx}^{\rm even}(z)$
but with the
mechanical force switched off
for layers Pt1 through Pt12.
Solid lines: Exponential fits according to
Eq.~\eqref{eq_trend_pt12_pt20_sot},
Eq.~\eqref{eq_trend_pt1_pt10_sot},
Eq.~\eqref{eq_trend_force_off_pt1_pt12_sot}
and
Eq.~\eqref{eq_trend_force_off_pt13_pt17_sot}.
}
\end{figure}

In Fig.~\ref{fig_even_torque_mechanical_force_on_off}
we show the linear response coefficients of the
layer-resolved spin current
${q}_{yx}^{\rm even}(L)$
as diamonds for
two values of
broadening, $\Gamma$~=~25~meV and $\Gamma$~=~100~meV (see
Eq.~\eqref{eq_layer_resolved_spin_current_sot} for the
definition of ${q}_{yx}^{\rm even}(L)$).
Evaluating the SHE-to-SOT conversion efficiency
defined in Eq.~\eqref{eq_estimate_tyx_from_she}
from the ratio of
torkance to maximal spin current we obtain
$\xi_{\rm 25 meV}^{\phantom{25}}=
{\torkance}_{yx}^{\rm even}/
[{q}_{yx}^{\rm even}(L={\rm Pt11})]$~=~0.74.
At $\Gamma$~=~100~meV the value is slightly lower:
$\xi_{\rm 100 meV}^{\phantom{100}}$~=~0.57. These values of $\xi$
resemble the experimentally determined spin-current 
transmissivities in Pt-based magnetic 
bilayer systems~\cite{comparison_sot_spin_pumping_nifept}.

Computing the electric conductivities based on the
same formalism as used for SOT and ISOT, we obtain
$\sigma_{xx}^{\rm 25meV}$~=~$1.26\cdot 10^7$~S/m
and
$\sigma_{xx}^{\rm 100meV}$~=~$0.34\cdot 10^7$~S/m.
From these conductivities
and the spin currents at the center of Pt, which are
given by ${q}_{yx}^{\rm even}(L={\rm Pt11})$,
we obtain the following SHE angles:
$\tan\gamma_{\rm SHE}^{\rm 25meV}$~=~0.029
and
$\tan\gamma_{\rm SHE}^{\rm 100meV}$~=~0.109.
The SHE angle increases thus by a factor of 3.8 as 
$\Gamma$ is increased from 25~meV to 100~meV. This
increase of the SHE angle with increasing disorder
is expected for the intrinsic SHE, because the
intrinsic SHE 
conductivity $\sigma_{zx}^{y}$ (see Eq.~\eqref{eq_she_current_model}) 
depends only weakly on disorder, while
the normal conductivity $\sigma_{xx}$ decreases with disorder. 
Indeed, the increase
of the SHE angle by the factor of 3.8 is well explained
by the ratio $\sigma_{xx}^{\rm 25meV}/\sigma_{xx}^{\rm 100meV}=3.7$.  

At $\Gamma$~=~100~meV the line
of blue diamonds illustrating ${q}_{yx}^{\rm even}(z)$ in
Fig.~\ref{fig_even_torque_mechanical_force_on_off}
is
constant in the central
region between Pt5 and Pt15, because
the primary spin current generated by SHE is constant in this
region and because secondary spin currents arising from the
reflections of spin current at the surfaces and interfaces decay 
strongly spatially 
and therefore do not reach
the central region between Pt5 and Pt15.
One reason for the suppression of the spin current-profile
in the region between Pt1 and Pt5 and in the region between Pt15
and Pt20 is the interference of the primary spin current from
the SHE with secondary spin current reflected respectively from the
surface and the interface. Additionally, as discussed in 
Sec.~\ref{sect_consequences_reciprocity},
we expect that the
primary spin current generated by the SHE is itself dependent on
position in these two regions and not constant like in the central region. 
In particular for the higher broadening of $\Gamma$~=~100~meV
the spin current profiles from our \textit{ab initio} calculations
shown in Fig.~\ref{fig_even_torque_mechanical_force_on_off} 
exhibit exponential behavior in the region Pt1 through Pt5 and
in the region Pt15 through Pt20. 
At $\Gamma$~=~100~meV the spin current in the region between
Pt12 and Co1 is
well described by the exponential fit
\bege\label{eq_trend_pt12_pt20_sot}
{q}_{yx}^{\rm even}(z)=
\left[
0.97
-0.35
e^{
(z-z_{\rm Pt20}^{\phantom{Pt}})/
\lambda_{\rm SOT,3}^{\rm 100 meV}
}
\right]ea_{0},
\ee
where $z_{\rm Pt20}^{\phantom{Pt}}$ is the $z$ coordinate of layer Pt20
and $\lambda_{\rm SOT,3}^{\rm 100 meV}$~=~0.46~nm.
In the region from Pt1 to Pt10
the spin current is approximately given by
\bege\label{eq_trend_pt1_pt10_sot}
{q}_{yx}^{\rm even}(z)=
\left[
0.92-0.63e^{-(z-z^{\phantom{Pt}}_{\rm Pt1})/\lambda_{\rm SOT,4}^{\rm 100 meV}}
\right]ea_{0}
\ee
with $\lambda_{\rm SOT,4}^{\rm 100 meV}$~=~0.32~nm.
At the smaller broadening of $\Gamma=25$~meV
we find
$\lambda_{\rm SOT,4}^{\rm 25 meV}=0.15$~nm,
but due to
oscillations the first principles data are less
well described by the
exponential fit.

The length $\lambda_{\rm SOT,4}^{\rm 100 meV}$
describes the decay of spin current close to
the vacuum boundary at Pt1, while the
length $\lambda_{\rm SOT,3}^{\rm 100 meV}$ describes
the decay of spin current close to the Co layer. 
In order to investigate whether $\lambda_{\rm SOT,4}^{\rm 100 meV}$
and $\lambda_{\rm SOT,3}^{\rm 100 meV}$ simply describe the decay
of secondary reflected spin current or whether they additionally
exhibit a modification due to a potential position-dependence of
the primary spin current, we 
divide the Pt layer into two regions:
In the atomic layers Pt1 through
Pt12 we switch off the mechanical force $-e\vn{E}$
that the electrons
would otherwise experience due to the applied
electric field $\vn{E}$.
Only the atomic layers Pt13 through Co3 are
subject to the mechanical
force $-e\vn{E}$ in this modified calculation, which
is based on Eq.~\eqref{eq_mech_force_off_region}.
Thus, only
Pt13 through Pt20 generate sizable
SHE spin current
(SHE in Co is small).
The corresponding
linear response coefficients are shown
in
Fig.~\ref{fig_even_torque_mechanical_force_on_off}
as circles for
two values of
broadening, $\Gamma$~=~25~meV
and $\Gamma$~=~100~meV.
Switching off the mechanical force significantly
perturbs the spin-current profile in the
region Pt1 through Pt14 while
from Pt15 onwards, the two spin-current
profiles merge.
Approaching the region with mechanical force
switched off, i.e., approaching Pt12,
the spin current
(red circles in
Fig.~\ref{fig_even_torque_mechanical_force_on_off})
in region Pt13 to Pt17 is suppressed according
to
\bege\label{eq_trend_force_off_pt13_pt17_sot}
{q}_{yx}^{\rm even}(z)=
\left[0.84
-
0.52
e^{
-(z-z^{\phantom{Pt1}}_{\rm Pt12})
/
\lambda^{\rm 100meV}_{\rm SOT,5}
}
\right]ea_{0},
\ee
where $\lambda^{\rm 100meV}_{\rm SOT,5}$~=~0.31~nm.
We find a slight $\Gamma$-dependence:
$\lambda^{\rm 25meV}_{\rm SOT,5}$~=~0.28~nm.
In the region from Pt1 through Pt12
the spin current is well described by
\bege\label{eq_trend_force_off_pt1_pt12_sot}
{q}_{yx}^{\rm even}(z)=0.33
e^{
(z-z^{\phantom{Pt1}}_{\rm Pt12})
/
\lambda^{\rm 100meV}_{\rm SOT,2}
}ea_{0},
\ee
with $\lambda^{\rm 100meV}_{\rm SOT,2}$~=~0.85~nm.
At $\Gamma$~=~25~meV the spin-current profile in the
region Pt1 through Pt14 cannot be described well by an 
exponential fit. 

Comparing the lengths obtained from the exponential
fits in 
Eq.~\eqref{eq_trend_pt12_pt20_sot},
Eq.~\eqref{eq_trend_pt1_pt10_sot},
Eq.~\eqref{eq_trend_force_off_pt1_pt12_sot}
and
Eq.~\eqref{eq_trend_force_off_pt13_pt17_sot} 
we find that $\lambda^{\rm 100meV}_{\rm SOT,2}$ is
substantially larger than the other three lengths:
$\lambda^{\rm 100meV}_{\rm SOT,2}>\lambda^{\rm 100meV}_{\rm SOT,3}\approx \lambda^{\rm 100meV}_{\rm SOT,4}\approx \lambda^{\rm 100meV}_{\rm SOT,5}$.
The length $\lambda^{\rm 100meV}_{\rm SOT,2}$ describes the decay of
spin current in Pt in a region of space where no spin current is
generated (because the mechanical force is switched off 
in the region
Pt1 through Pt12). This spin current, which is injected
into the region Pt1-Pt12, originates only from
the SHE in the region Pt13-Pt20. 
In contrast, the lengths $\lambda^{\rm 100meV}_{\rm SOT,3}$,
$\lambda^{\rm 100meV}_{\rm SOT,4}$ and $\lambda^{\rm 100meV}_{\rm SOT,5}$
describe the suppression of the total spin current close to
interfaces and surfaces. The total spin current is the sum
of spin current generated by SHE and spin current from the
reflection at interfaces and surfaces. This reflected spin current
is expected to be described by $\lambda^{\rm 100meV}_{\rm SOT,2}$.
Our finding that $\lambda^{\rm 100meV}_{\rm SOT,3}$,
$\lambda^{\rm 100meV}_{\rm SOT,4}$ and $\lambda^{\rm 100meV}_{\rm SOT,5}$
are all much smaller than $\lambda^{\rm 100meV}_{\rm SOT,2}$ can only
be explained if we assume that the primary spin current generated by
SHE is itself modified close to surfaces and interfaces.

\begin{figure}
\includegraphics*[width=8.5cm]{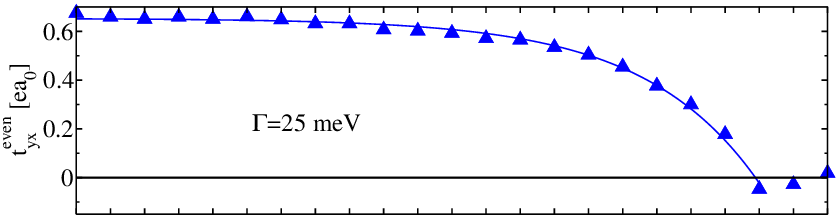}
\includegraphics*[width=8.7cm]{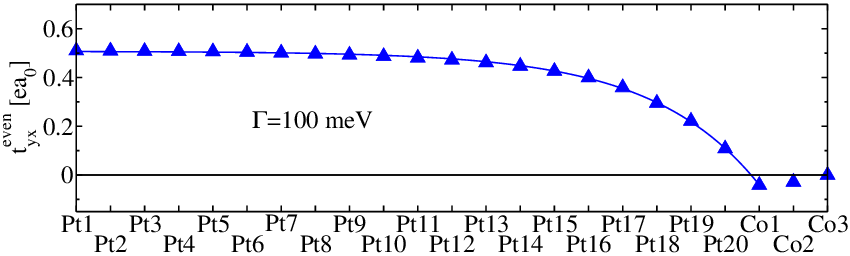}
\caption{\label{fig_mech_force_off_layerwise}
Triangles: Torkances
for broadenings of
$\Gamma=25$~meV (upper graph)
and  $\Gamma=100$~meV (lower graph).
For a given layer $L$ ($L$ is specified on the horizontal axis),
the mechanical force is switched off in the region
from Pt1 through $L-1$ according to
Eq.~\eqref{eq_mech_force_off_region} and the resulting
total torkance is shown by a blue triangle.
Solid lines: Exponential fits according to
Eq.~\eqref{eq_mechanical_force_off_layerwise}.
}
\end{figure}

In Fig.~\ref{fig_mech_force_off_layerwise} we show
the torkance as a function of
the region where the mechanical
force is set to zero.
If the
mechanical force is switched off in all Pt layers and only active
in the Co layers (data points at $L$=Co1),
${\torkance}_{yx}^{\rm even}$ is very small because
the even torque arises dominantly from the SHE in Pt which is switched off
when the mechanical force is set to zero.
When the mechanical force is set to zero
in the region from Pt1 through layer $L-1$, the torkance
is well described by the fit
\bege\label{eq_mechanical_force_off_layerwise}
{\torkance}_{yx}^{\rm even}(z)=
\left[
0.65
-0.68
e^{-\left(
z^{\phantom{Co1}}_{\rm Co1}-z
\right)/
\lambda^{\rm 25 meV}_{\rm SOT,1}}
\right]ea_{0},
\ee
where 
$\lambda^{\rm 25 meV}_{\rm SOT,1}$~=~0.76~nm.
We find a weak $\Gamma$-dependence:
$\lambda^{\rm 100 meV}_{\rm SOT,1}$~=~0.71~nm.
At $\Gamma=100$~meV the spin current generated in a 
given atomic layer of Pt decays
on the length scale of $\lambda^{\rm 100 meV}_{\rm SOT,2}$. Therefore,
the SHE from layers $L$ that are further away from the Co layer
than $\lambda^{\rm 100 meV}_{\rm SOT,2}$ cannot contribute 
to ${\torkance}_{yx}^{\rm even}$. Thus, we expect
$\lambda^{\rm 100 meV}_{\rm SOT,1}\approx\lambda^{\rm 100 meV}_{\rm SOT,2}$,
which is indeed the case.

One main conclusion of this subsection is that
for a sufficiently large broadening $\Gamma=100$~meV 
the \textit{ab initio} spin current profiles behave as expected
from diffusive spin transport models. In particular, 
at $\Gamma=100$~meV the decay lengths of spin current 
extracted in various ways are found to be similar, namely
$\lambda^{\rm 100 meV}_{\rm SOT,1}$~=~0.71~nm and 
$\lambda^{\rm 100meV}_{\rm SOT,2}$~=~0.85~nm. 
Similarly short but slightly longer length scales
of roughly 1.5~nm have been observed in Pt
in recent
experiments~\cite{current_induced_switching_using_spin_torque_from_spin_hall_buhrman,quantitative_study_spin_Hall_magnetoresistance,experimental_test_spin_mixing,determination_Pt_spin_diffusion_length_spin_pumping}.
A second conclusion from this subsection is that
close to interfaces and surfaces the SHE 
conductivity is position-dependent. Therefore, close to interfaces
and surfaces, the spin current
profiles do not decay on the scale 
of $\lambda^{\rm 100 meV}_{\rm SOT,1}\approx\lambda^{\rm 100 meV}_{\rm SOT,2}$ 
but instead significantly faster, namely according to
$\lambda^{\rm 100meV}_{\rm SOT,3}\approx \lambda^{\rm 100meV}_{\rm SOT,4}\approx \lambda^{\rm 100meV}_{\rm SOT,5}\approx 0.3{\rm nm}$.

\subsection{Even ISOT}
\label{sec_even_isot}
\begin{figure}
\includegraphics[width=8.65cm]{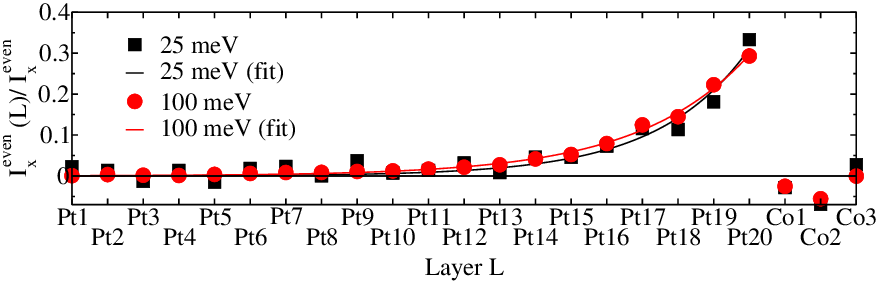}
\caption{\label{fig_isot_current_distribution} 
Layer-resolved
ISOT current $I_{x}^{\,\rm even}(L)$
induced
in Co(3)/Pt(20) by magnetization dynamics.
The total ISOT current is
$I_{x}^{\,\rm even}=\sum_{L}I_{x}^{\,\rm even}(L)$.
The relative contributions of the layers, i.e., 
$I_{x}^{\,\rm even}(L)/I_{x}^{\,\rm even}$, is shown
for two values of broadening, $\Gamma$~=~25~meV and
$\Gamma$~=~100~meV.
Solid lines: Exponential fit according 
to Eq.~\eqref{eq_decay_even_charge_current_pumped}.
}
\end{figure}
When the magnetization precesses in a circular orbit
around the $z$ axis 
in the small-cone limit the 
current density
\bege\label{eq_pumped_current_CoPt}
\begin{aligned}
\frac{J^{\,\rm even}_{x, \rm\,25meV}(t)}{\omega}=&
-87\frac{\rm pAs}{\rm m}
\sin(\theta)\sin(\omega t),
\\
\frac{J^{\,\rm even}_{x, \rm\,100meV}(t)}{\omega}=&
-68\frac{\rm pAs}{\rm m}
\sin(\theta)\sin(\omega t)
\end{aligned}
\ee
is induced due to the even torkance 
$t_{yx}^{\rm even}$ according 
to Eq.~\eqref{eq_induced_currents_z}, 
where we used $A_{0}=t_{yx}^{\rm even}$ and
$A=23.8\,a_0^2$.

As discussed in 
Sec.~\ref{sect_consequences_reciprocity},
the ISOT 
current $I^{\rm\,even}_{x}=J^{\rm\,even}_{x}L^{\phantom{e}}_{y}$ arises 
dominantly from the
combination of spin pumping and ISHE. 
Since the spin current pumped into Pt decays, 
the layer-resolved
ISOT current $I_{x}^{\rm even}(L)$ is expected to reflect this
spatial decay. 
Replacing $v_{\alpha}$ in 
Eq.~\eqref{eq_kubo_linear_response_torkance} by
$v_{\alpha}(L)$ (Eq.~\eqref{eq_modified_velocity_operator}) 
yields the layer-resolved ISOT current $I_{x}^{\rm even}(L)$ 
shown in Fig.~\ref{fig_isot_current_distribution}.
Inside the Pt layer, $I_{x}^{\rm even}(L)$
is well described
by an exponential function,
\bege\label{eq_decay_even_charge_current_pumped}
I_{x}^{\,\rm even}(z)
=I_{x}^{\,\rm even}(z^{\phantom{Pt20}}_{\rm Pt20})
e^{(z-z^{\phantom{Pt20}}_{\rm Pt20})/
\lambda_{\rm ISOT, 1}^{\phantom{\rm 25 meV}}},
\ee
where $z^{\phantom{Pt20}}_{\rm Pt20}$ is the $z$ coordinate 
of layer Pt20. 
Fitting Eq.~\eqref{eq_decay_even_charge_current_pumped} 
to the
$I_{x}^{\,\rm even}(L)$ profile
obtained from first principles 
yields $\lambda_{\rm ISOT, 1}^{\rm 25 meV}$~=~0.58~nm and
$\lambda_{\rm ISOT, 1}^{\rm 100 meV}$~=~0.70~nm.
 
\begin{figure}
\includegraphics[width=8.5cm]{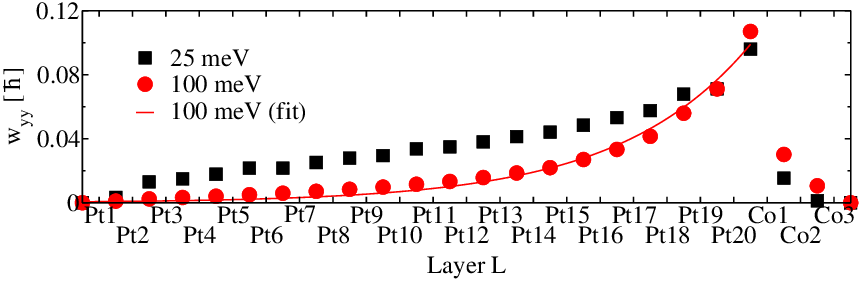}
\caption{\label{fig_isot_spin_current_distribution}
Layer-resolved
spin current induced by magnetization dynamics
for two values of broadening, $\Gamma$~=~25~meV and
$\Gamma$~=~100~meV.
The coefficient $w_{yy}$ describes spin current flowing
in $z$ direction with spin pointing in $y$ direction and
in phase with $I_{x}^{\,\rm even}(t)$.
Solid line: Fit according to
Eq.~\eqref{eq_isot_spin_current_profile_even}.
}
\end{figure}

In order to compare 
the spatial profile of the
layer-resolved ISOT current $I_{x}^{\rm even}(L)$ with the
spatial profile of the pumped spin current ${Q}_{y}(L,t)$
given by Eq.~\eqref{eq_induced_spin_current}, we calculate
the coefficients $w^{\phantom{(a)}}_{yy}(L)$, which are defined 
in Eq.~\eqref{eq_kubo_linear_response_pumped_spin_current}.
$w^{\phantom{(a)}}_{yy}(L)$
describes spin current in phase with $I_{x}^{\,\rm even}$
and with spin pointing in $y$ direction.
Within Pt, the $L$-dependence of $w^{\phantom{(a)}}_{yy}(L)$, 
shown 
in Fig.~\ref{fig_isot_spin_current_distribution}, is
approximately given by
\bege\label{eq_isot_spin_current_profile_even}
w^{\phantom{(a)}}_{yy}(z)
=0.087\hbar
e^{
\left(
z
-
z_{\rm Pt20}^{\phantom{Pt1}}
\right)
/\lambda^{\rm 100meV}_{\rm ISOT,2}
},
\ee
where $\lambda^{\rm 100meV}_{\rm ISOT,2}$~=~0.89~nm.
At smaller broadening $\Gamma$~=~25~meV, 
the pumped spin current reaches the vacuum boundary
at Pt1 and the resulting reflection of spin current
needs to be considered according to
Eq.~\eqref{eq_spin_current_profile_extended}. 
When $\lambda^{\rm 25meV}_{\rm ISOT,2}$ is much 
larger than the thickness of Pt, the
$\sinh$ function can be approximated:
\bege\label{eq_linear_spin_current_profile}
w^{\phantom{(a)}}_{yy}(z)
\propto
\frac{\sinh{\frac{z-z_{\rm Pt1}^{\phantom{P}}}{ \lambda^{\rm 25meV}_{\rm ISOT,2} }}}
{\sinh{\frac{ z_{\rm Pt20}^{\phantom{P}}-z_{\rm Pt1}^{\phantom{P}}
    }{ \lambda^{\rm 25meV}_{\rm ISOT,2}   }}}
\approx
\frac{
z-z_{\rm Pt1}^{\phantom{P}}
}{
z_{\rm Pt20}^{\phantom{P}}-z_{\rm Pt1}^{\phantom{P}}
}
,
\ee
which explains the roughly linear profile of
$w^{\phantom{(a)}}_{yy}(L)$ at $\Gamma$~=~25~meV.

The ISOT currents shown in 
Fig.~\ref{fig_isot_current_distribution} decay 
faster in Pt 
than the spin currents in Fig.~\ref{fig_isot_spin_current_distribution}. Thus,
Eq.~\eqref{eq_position_dependence_diffusive_model},
which predicts spin current and ISHE-current to be
proportional, is violated, in 
particular at
$\Gamma$~=~25~meV. However, 
Eq.~\eqref{eq_position_dependence_diffusive_model} is 
approximately
satisfied
at $\Gamma$~=~100~meV, where both the ISOT current and the
pumped spin current decay exponentially 
with $\lambda^{\rm 100meV}_{\rm ISOT,1}\approx \lambda^{\rm 100meV}_{\rm ISOT,2}$.
The small difference 
$\lambda^{\rm 100meV}_{\rm ISOT,2}-\lambda^{\rm 100meV}_{\rm ISOT,1}$~=~0.19~nm 
amounts to less than one Pt interlayer 
distance. Additionally, this spin current decay 
length $\lambda^{\rm 100meV}_{\rm ISOT,1}\approx \lambda^{\rm 100meV}_{\rm ISOT,2}$
is very similar to the one extracted in the previous subsection, i.e.,
$\lambda^{\rm 100meV}_{\rm ISOT,1}\approx \lambda^{\rm 100meV}_{\rm ISOT,2}
\approx\lambda^{\rm 100meV}_{\rm SOT,1}\approx \lambda^{\rm 100meV}_{\rm SOT,2}$.
This consistency between the various methods used to extract the 
spin current decay length implies that the  
model of Sec.~\ref{sect_consequences_reciprocity} provides
a satisfactory description at sufficiently high broadening $\Gamma$. 

In Eq.~\eqref{eq_charge_current_due_to_ishe_model}
the ISHE angle $\tan\gamma_{\rm ISHE}^{\phantom{I}}$ is proportional to the
quotient of ISOT current density and pumped spin current density.
The different decay of ISOT current 
and pumped spin current described 
by $\lambda^{\rm 100meV}_{\rm ISOT,1}$
and $\lambda^{\rm 100meV}_{\rm ISOT,2}$, respectively,
therefore implies that $\tan\gamma_{\rm ISHE}^{\phantom{I}}$ 
is not constant but
dependent on position.
For large broadening we obtain
$\tan\gamma_{\rm ISHE}^{\rm 100meV}(L={\rm Pt20})$~=~0.16
and
$\tan\gamma_{\rm ISHE}^{\rm 100meV}(L={\rm Pt11})$~=~0.077,
while for small broadening we obtain
$\tan\gamma_{\rm ISHE}^{\rm 25meV}(L={\rm Pt20})$~=~0.27
and
$\tan\gamma_{\rm ISHE}^{\rm 25meV}(L={\rm Pt11})$~=~0.031.
Even for large broadening, the ISHE angle is significantly
enhanced at the interface. 
The ISHE angles at the centre of Pt, 
i.e., $\tan\gamma_{\rm ISHE}^{\rm 100meV}(L={\rm Pt11})$
and $\tan\gamma_{\rm ISHE}^{\rm 25meV}(L={\rm Pt11})$,
are similar to the SHE angles determined in the previous section
from the spin current in the center of
Pt: $\tan\gamma_{\rm SHE}^{\rm 100meV}(L={\rm Pt11})$~=~0.109 
and $\tan\gamma_{\rm SHE}^{\rm 25meV}(L={\rm Pt11})$~=~0.029.

From Eq.~\eqref{eq_mix_fp} we obtain 
the spin-mixing conductance ${\rm Re}g^{\uparrow\downarrow}_{\rm 25meV}
=1.8\cdot 10^{19}$~m$^{-2}$
and for $\Gamma$~=~100~meV
a slightly larger value of
${\rm Re}g^{\uparrow\downarrow}_{\rm 100meV}
=2.0\cdot 10^{19}$~m$^{-2}$. 
Eq.~\eqref{eq_lambda_ishe_connection} provides an alternative
way to extract the ISHE angle: 
\bege\label{eq_lambda_ishe_connection_testit}
\tan\bar{\gamma}_{\rm ISHE}^{\rm 100meV}
=
\frac{
2\pi t^{\rm even}_{yx}
}
{
e A {\rm Re}
g^{\uparrow\downarrow}_{\rm 100meV}
\lambda^{\rm 100meV}_{\rm ISOT,2}
}=0.15,
\ee
where the in-plane area of the unit cell is $A=23.8\,a_0^2$
and the parameters $\lambda^{\rm 100meV}_{\rm ISOT,2}$~=~0.89~nm
and ${\torkance}_{yx,\rm\,100meV}^{\rm even}=0.53\,ea_{0}$ have been discussed
above. In contrast to the layer-resolved ISHE 
angles, Eq.~\eqref{eq_lambda_ishe_connection_testit} describes an average
over all those Pt layers that lie within the 
distance of $\lambda^{\rm 100meV}_{\rm ISOT,2}$ from the Co layer.
The result of $\tan\bar{\gamma}_{\rm ISHE}^{\rm 100meV}=0.15$ is very similar to the
layer-resolved ISHE angle close to the interface 
of $\tan\gamma_{\rm ISHE}^{\rm 100meV}(L={\rm Pt20})$~=~0.16.

Finally, we can also put Eq.~\eqref{eq_intrinsic_lambda} to a test 
using the parameters
determined above:
\bege\label{eq_intrinsic_lambda_testit}
\lambda_{\rm sd}^{100\rm meV}=\frac{\xi_{100\rm meV}
\hbar \pi\sigma_{xx}^{100\rm meV}}{e^2 {\rm Re}
g^{\uparrow\downarrow}_{100\rm meV}}=1.25{\rm nm}.
\ee
While $\lambda_{\rm sd}^{100\rm meV}$ is 
larger than $\lambda^{\rm 100meV}_{\rm ISOT,2}$, the agreement between
these two values is still
satisfactory, corroborating the conclusion that the model
of Sec.~\ref{sect_consequences_reciprocity} provides a satisfactory description 
for sufficiently large broadening.
For small broadening Eq.~\eqref{eq_intrinsic_lambda} 
yields $\lambda_{\rm sd}^{25\rm meV}=6.7{\rm nm}$,
which is thicker than the Pt layer in our calculation and therefore justifies
the
linear approximation in Eq.~\eqref{eq_linear_spin_current_profile}.

\subsection{Odd SOT}
\label{sec_odd_sot}
We obtain 
torkances per unit cell of
${\torkance}_{xx,{\rm 25meV}}^{\rm odd}=0.17\,ea_{0}$ 
and
${\torkance}_{xx,{\rm 100meV}}^{\rm odd}=0.15\,ea_{0}$
at broadenings of
$\Gamma$~=~25~meV 
and $\Gamma$~=~100~meV, respectively.  
Dividing these torkances by the magnetic
moment per unit cell
of $\mu=$5.78~$\mu_{\rm B}$ we calculate the
effective fields per applied
electric field
of
${\torkance}_{xx,{\rm 25meV}}^{\rm odd}/\mu=0.0027\,{\rm mTcm/V}$
and
${\torkance}_{xx,{\rm 100meV}}^{\rm odd}/\mu=0.0024\,{\rm mTcm/V}$.

\begin{figure}
\includegraphics*[width=8.5cm]{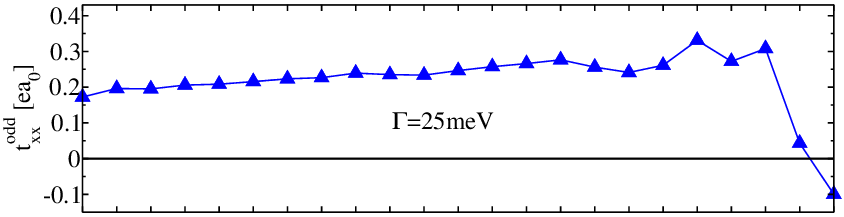}
\includegraphics*[width=8.7cm]{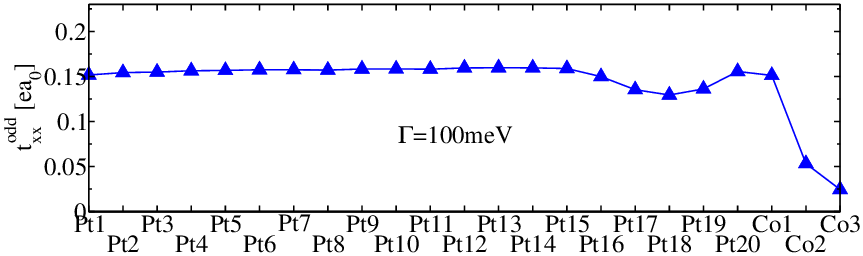}
\caption{\label{fig_mech_force_off_layerwise_odd} 
Triangles: Torkances 
for broadenings of 
$\Gamma=25$~meV (upper graph) 
and  $\Gamma=100$~meV (lower graph). 
For a given layer $L$ ($L$ is specified on the horizontal axis), 
the mechanical force is switched off in the region
from Pt1 through $L-1$ according to
Eq.~\eqref{eq_mech_force_off_region} and the resulting total
torkance is shown by a blue triangle.
Solid lines serve as guide to the eye.
}
\end{figure}

In Fig.~\ref{fig_mech_force_off_layerwise_odd} we show 
the odd torkance as a function of the region
with mechanical force switched off.
If the mechanical force is switched off 
for Pt1 through Pt20 such that only the layers
Co1, Co2 and Co3 are subject to it
(see the data 
points at $L$=Co1 in the figure), the corresponding odd
torque is not very different from the one with the
mechanical force switched on everywhere (see the data
points at $L$=Pt1 in the figure).
If the mechanical force is applied only to layers Co2 and Co3
(see data points at $L$=Co2 in the figure) the resulting torkance
is much smaller compared to the situation where all three Co
layers are subject to it. Thus, the perturbation of the Co1 layer
by the mechanical force is essential for the odd SOT in this system.

To produce a sizable odd torque in Co(3)/Pt(20)
it is therefore not crucial to switch on the mechanical
force in the Pt layers but it suffices to apply this
perturbation to the Co states. 
As a combined effect of broken inversion symmetry
and SOI the
spin of a given
wave function $|\Psi_{\vn{k}n}\rangle$ is correlated with the
velocity $v_{\vn{k}nn}$~\cite{edelstein}. 
As a result, the non-equilibrium
spin density induced by an applied electric field
combined with the exchange interaction gives rise
to the odd component of the 
torkance~\cite{torque_macdonald,chernyshov_2009,manchon_zhang_2009}. 
Application of the mechanical force to Co,
i.e., perturbation of the system via the velocity operator
within the Co layer, produces therefore the dominant
part of nonequilibrium spin density from which the
odd torque arises in Co(3)/Pt(20).
This stands 
in marked contrast
to the even torque in this system, which
is mainly driven by SHE from Pt and thus very small if the
mechanical force is turned off in all Pt layers, as
shown in Fig.~\ref{fig_mech_force_off_layerwise}. 

\begin{figure}
\includegraphics*[width=8.5cm]{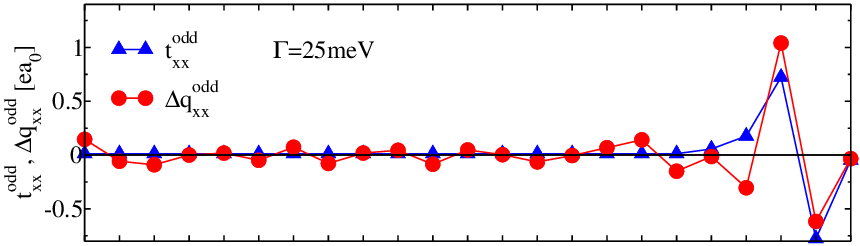}
\includegraphics*[width=8.7cm]{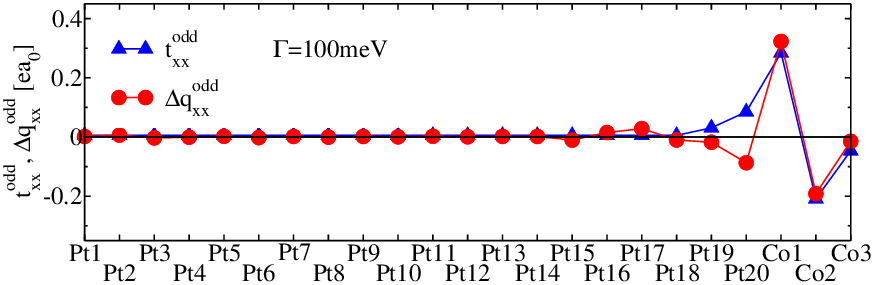}
\caption{\label{fig_layer_resolved_odd_torkance}
Triangles: Layer-resolved odd
torkance ${\torkance}_{xx}^{\rm odd}$
for broadenings of
$\Gamma=25$~meV (upper graph)
and  $\Gamma=100$~meV (lower graph).
Circles: Linear response coefficient of the
layer-resolved spin flux $\Delta q_{xx}^{\rm odd}$,
Eq.~\eqref{eq_spin_current_flux_odd}.
Solid lines serve as guide to the eye.
}
\end{figure}

In Fig.~\eqref{fig_layer_resolved_odd_torkance} the
layer-resolved odd torkance and the linear-response
coefficient of spin flux into layer $L$, i.e.,
\bege\label{eq_spin_current_flux_odd}
\Delta q_{xx}^{\rm odd}(L)=
q_{xx}^{\rm odd}(L)-
q_{xx}^{\rm odd}(L+1),
\ee
are shown for two values of broadening, $\Gamma$~=~25~meV
and $\Gamma$~=~100~meV. For the layers Co1 through Co3
the layer-resolved torkances coincide approximately with the
spin fluxes like in the case of the even torque. 
This approximate agreement between odd spin fluxes and odd torques is
not generally found in bilayer systems, for example they
differ considerably in O/Co/Pt and Al/Co/Pt~\cite{ibcsoit}.
For 
$\Gamma$~=~100~meV the magnetization of layer Pt20 
experiences a torkance of 0.085$ea_0$. At the same time
there is a spin flux out of layer Pt20 characterized by the
coefficient $-\Delta q_{xx}^{\rm odd}(L={\rm Pt20})=0.087ea_0$. 
This spin flux is transferred to the Co layer 
where it exerts a torque
on the Co magnetization. The sum of torkance and spin flux
coefficient of Pt20 amounts 
to 0.172$ea_0$ and approximately accounts for
the total odd torkance of 0.15$ea_0$ at $\Gamma$~=~100~meV.
The angular momentum that gives rise to the odd torque on the
magnetization is thus picked up from the lattice
at Pt20 and roughly 50\% of it is directly transferred to the
magnetization of the Pt20 layer while the rest is transported
to the Co layer via spin current.
Above we have shown that the mechanical force on the Co1 layer
is crucial to produce a sizable odd torque.
Since the pick-up of angular momentum from the lattice by 
the spin system happens in Pt20, the hybridization of the Co1
states with the Pt20 states is thus essential. 

\subsection{Odd ISOT}
\label{sec_odd_isot}

According to Eq.~\eqref{eq_induced_currents_z}
the current density
\bege\label{eq_pumped_current_odd_CoPt}
\begin{aligned}
\frac{J^{\,\rm odd}_{x,{\rm 25meV}}(t)}{\omega}=&
22\frac{\rm pAs}{\rm m}
\sin(\theta)\cos(\omega t),
\\
\frac{J^{\,\rm odd}_{x,{\rm 100meV}}(t)}{\omega}=&
19\frac{\rm pAs}{\rm m}
\sin(\theta)\cos(\omega t)
\end{aligned}
\ee
is induced due to ${\torkance}_{xx}^{\rm odd}$
when the magnetization precesses around the $z$ axis
in the small-cone limit.
Here, we used
$C_{0}=t_{xx}^{\rm odd}$ and $A=23.8\,a_0^2$. This
contribution
from ${\torkance}_{xx}^{\rm odd}$ is thus $-90^{\circ}$ phase
shifted
with respect to the contribution from ${\torkance}_{yx}^{\rm even}$
given in Eq.~\eqref{eq_pumped_current_CoPt}, i.e., it lags
behind by a quarter period.

\begin{figure}
\includegraphics*[width=8.7cm]{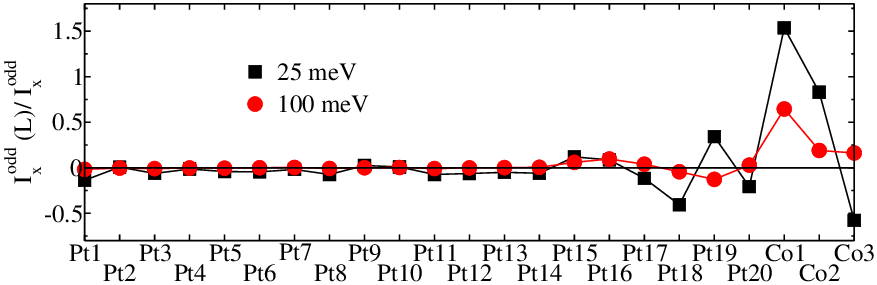}
\caption{\label{fig_isot_current_odd}
Layer-resolved
ISOT current $I_{x}^{\,\rm odd}(L)$
induced
in Co(3)/Pt(20) by magnetization dynamics.
The total ISOT current is
$I_{x}^{\,\rm odd}=\sum_{L}I_{x}^{\,\rm odd}(L)$.
The relative contributions of the layers, i.e.,
$I_{x}^{\,\rm odd}(L)/I_{x}^{\,\rm odd}$, is shown
for two values of broadening,
$\Gamma$~=~25~meV (squares) and
$\Gamma$~=~100~meV (circles).
Solid lines serve as guide to the eye.
}
\end{figure}

Since the mechanical force on the Co1 layer is
crucial for the odd SOT according to 
Fig.~\ref{fig_mech_force_off_layerwise_odd}
we expect
that the odd ISOT current induced by
magnetization dynamics flows mainly in the Co1 layer, because
of the reciprocity between ISOT and SOT. 
This
is indeed the case, as Fig.~\ref{fig_isot_current_odd} shows. In
particular, at $\Gamma$~=~100~meV
the currents flowing in Co2, Co3 and the Pt layer are almost
negligible. At the smaller broadening $\Gamma$~=~25~meV
the induced ISOT currents in Co2, Co3 and Pt are larger,
especially in the Co2 and Co3 layers, but the Co1 contribution
to the ISOT current still strongly dominates.

\section{Summary}
\label{sec_summary}
SOT and ISOT are reciprocal effects.
Both of them
can be expressed conveniently in terms of 
the torkance tensor $\vn{t}(\hat{\vn{M}})$,
which depends on the 
magnetization direction $\hat{\vn{M}}$.
In the case of the SOT phenomenon,
the torque $\vn{T}(\hat{\vn{M}})$ on the 
magnetization due to the application of an
electric field $\vn{E}$ is 
given by $\vn{T}(\hat{\vn{M}})=\vn{t}(\hat{\vn{M}})\vn{E}$. If $\hat{\vn{M}}$  changes
as a function of time, the reciprocal effect, 
the ISOT, can be observed.
It consists in the generation of a current 
density $\vn{j}(t)=[\vn{t}(-\hat{\vn{M}}(t))]^{\rm T}
[\hat{\vn{M}}(t)\times\frac{d \hat{\vn{M}}(t)}{dt}]/V$, 
where $V$ is the unit cell volume.
Magentization dynamics driven effects, such as ISOT and
Gilbert damping, can be consistently derived in time-dependent
perturbation theory using a time-dependent exchange field.
The same expressions are obtained by rewriting general
many body susceptibilities in terms of the Kohn-Sham 
susceptibilities.
On the basis of the SOT-ISOT reciprocity relations and 
recent experimental results for the SOT in 
bilayer systems, we predict the angular 
dependence of the FMR-driven ISOT in
bilayers. We find that measurements of 
the dc voltage associated with the 
FMR-driven ISOT are insufficient 
to determine $\vn{t}(\hat{\vn{M}})$ in general
and that additionally the ac voltage needs to be 
measured phase-sensitively to 
determine $\vn{t}(\hat{\vn{M}})$ completely.   
Within the Kubo linear response formalism we investigate
SOTs and ISOTs in Co/Pt(111) magnetic bilayers using the electronic
structure provided from first principles density functional theory.
Magnetization-dynamics induced charge currents and spin currents are resolved
on the atomic scale to extract model parameters and to expose
the mechanisms underlying the ISOT. Likewise the
spin currents accompanying the SOT are resolved on the atomic scale for the
same purposes. It is found that SHE and ISHE are modified 
close to interfaces and surfaces. Comparison of the various currents
accompanying SOT on the one hand and ISOT on the other hand highlights
the reciprocity of the two phenomena on the microscopic scale.  

\acknowledgments
We gratefully acknowledge computing time on the supercomputers \mbox{JUQUEEN}
and \mbox{JUROPA}
at J\"ulich Supercomputing Center and funding under the HGF-YIG programme VH-NG-513.

\appendix
\section{Magnetocrystalline anisotropy and the static torque-torque correlation function}
\label{app_mae}
The torque due to the 
field $\vn{H}^{\rm MAE}$, Eq.~\eqref{eq_internal_mag_field}, is given by
\begin{gather}\label{eq_mae1}
\begin{aligned}
\delta\vn{T}^{\rm MAE}&=\mu_{0}MV\hat{\vn{M}}\times\vn{H}^{\rm MAE}\\
&=-MV
\left[
\bar{\Omega}^{\rm xc}
+
\frac{
G^{\rm R}_{\vn{\mathcal{T}}\vn{\mathcal{T}}}
(\hbar\omega=0,\hat{\vn{M}})
}{MV\hbar}
\right]
\hat{\vn{M}}\times\delta \hat{\vn{M}},
\end{aligned}\raisetag{3\baselineskip}
\end{gather}
where we used Eq.~\eqref{eq_chi_final} to express 
$\vn{\chi}$ in terms of the torque-torque 
correlation function $G^{\rm R}_{\vn{\mathcal{T}}\vn{\mathcal{T}}}$. 
Eq.~\eqref{eq_mae1} 
can be related easily to anisotropy constants.
For example, in the case of uniaxial anisotropy, 
i.e., $E(\theta)=VK_1\sin^2\theta$, one obtains
\bege
K_{1}=\frac{M}{2}
\left[
\bar{\Omega}^{\rm xc}
+
\frac{
G^{\rm R}_{\mathcal{T}_{y}\mathcal{T}_{y}}
(\hbar\omega=0,\hat{\vn{M}}=\hat{\vn{e}}_{z})
}{MV\hbar}
\right].
\ee

In the following we show that Eq.~\eqref{eq_mae1}, which was 
obtained within the many-electron
response formalism of Sec.~\ref{sec_many_electron_resp}, 
can also be obtained
directly from the torque exerted on the magnetization by 
the Kohn-Sham electrons.
Denoting the Kohn-Sham wavefunctions by $|\psi_{\vn{k}n}\rangle$ and 
the occupancies by $f_{\vn{k}n}$
we can write
\bege
\label{eq_mae2}
\begin{aligned}
\delta\vn{T}^{\rm MAE}=&-\delta
\left\{
\frac{1}{\mathcal{N}}
\sum_{\vn{k}n}f_{\vn{k}n}
\langle \psi_{\vn{k}n} | \vht{\mathcal{T}}|\psi_{\vn{k}n}\rangle
\right\}\\
=&-\frac{1}{\mathcal{N}}
\sum_{\vn{k}n}
f_{\vn{k}n}\langle \psi_{\vn{k}n} | \delta\vht{\mathcal{T}}|\psi_{\vn{k}n}\rangle\\
&-\frac{1}{\mathcal{N}}
\sum_{\vn{k}n}
\delta f_{\vn{k}n}\langle \psi_{\vn{k}n} |\vht{\mathcal{T}}|\psi_{\vn{k}n}\rangle\\
&-2{\rm Re}\frac{1}{\mathcal{N}}
\sum_{\vn{k}n}
f_{\vn{k}n}\langle \psi_{\vn{k}n} |\vht{\mathcal{T}}\delta|\psi_{\vn{k}n}\rangle.
\end{aligned}
\ee
From $\delta \vht{\mathcal{T}}=\vn{m}\times \delta \hat{\vn{M}}\Bxc^{\rm xc}$ we obtain for
the first term
\bege
-\frac{1}{\mathcal{N}}
\sum_{\vn{k}n}
f_{\vn{k}n}\langle \psi_{\vn{k}n} | \delta\vht{\mathcal{T}}|\psi_{\vn{k}n}\rangle
=-MV\bar{\Omega}^{\rm xc}\hat{\vn{M}}\times \delta \hat{\vn{M}}.
\ee
Using for the remaining terms
\bege
\delta|\psi_{\vn{k}n}\rangle=\sum_{m\neq n}
\frac{|\psi_{\vn{k}m}\rangle\langle \psi_{\vn{k}m} |\vht{\mathcal{T}}|\psi_{\vn{k}n}\rangle}
{\mathcal{E}_{\vn{k}n}-\mathcal{E}_{\vn{k}m}}\cdot(\hat{\vn{M}}\times \delta \hat{\vn{M}})
\ee
and
\bege
\delta f_{\vn{k}n}=-\delta(\mathcal{E}_{\rm F}-\mathcal{E}_{\vn{k}n})
\langle \psi_{\vn{k}n} |\vht{\mathcal{T}}|\psi_{\vn{k}n}\rangle
\cdot
(\hat{\vn{M}}\times \delta \hat{\vn{M}})
\ee
and 
\begin{equation}\label{eq_retarded_torque_torque}
\begin{aligned}
G^{\rm R}_{\mathcal{T}_{\alpha}\mathcal{T}_{\beta}}&=
\frac{2\hbar}{\mathcal{N}}
\sum_{\vn{k}n}
\sum_{m\neq n}
f_{\vn{k}n}{\rm Re}
\frac{
\langle \psi_{\vn{k}n} |\mathcal{T}_{\alpha}|\psi_{\vn{k}m}\rangle
\langle \psi_{\vn{k}m} |\mathcal{T}_{\beta}|\psi_{\vn{k}n}\rangle
}{\mathcal{E}_{\vn{k}n}-\mathcal{E}_{\vn{k}m}}\\
&\!\!\!\!\!\!\!\!-\frac{\hbar}{\mathcal{N}}
\sum_{\vn{k}n}\delta(\mathcal{E}_{\rm F}-\mathcal{E}_{\vn{k}n})
\langle \psi_{\vn{k}n} |\mathcal{T}_{\alpha}|\psi_{\vn{k}n}\rangle
\langle \psi_{\vn{k}n} |\mathcal{T}_{\beta}|\psi_{\vn{k}n}\rangle
\end{aligned}
\end{equation}
one can easily show that Eq.~\eqref{eq_mae1} and Eq.~\eqref{eq_mae2} agree.

The Kohn-Sham Hamiltonian can be decomposed as
\bege
H(\vn{r})=H_{\rm KIN}+V(\vn{r})-\vn{m}\cdot\hat{\vn{M}}\Omega^{\rm xc}(\vn{r})+H_{\rm SOI},
\ee
where $H_{\rm KIN}$ describes the kinetic energy, $V(\vn{r})$ is the spin-independent part of the
effective potential and $H_{\rm SOI}$ describes the spin-orbit interaction. Using
$[H_{\rm KIN},\sigma_{\beta}]=0$, $[V(\vn{r}),\sigma_{\beta}]=0$ and $[\sigma_{\alpha},\sigma_{\beta}]=2i\epsilon_{\alpha\beta\gamma}\sigma_{\gamma}$
one can show the following identity for the torque operator:
\bege\label{eq_torque_soi_comm}
\mathcal{T}_{\beta}=\frac{i}{2}
\left[
H-H_{\rm SOI},\sigma_{\beta}
\right].
\ee
Substituting $\mathcal{T}_{\beta}$ in Eq.~\eqref{eq_retarded_torque_torque} 
by Eq.~\eqref{eq_torque_soi_comm} and inserting the 
resulting expression for $G^{\rm R}_{\mathcal{T}_{\alpha}\mathcal{T}_{\beta}}$
into Eq.~\eqref{eq_mae1} we obtain
\begin{gather}\label{eq_tmae_final}
\begin{aligned}
&\delta\vn{T}^{\rm MAE}=-
\frac{1}{\mathcal{N}}
\sum_{\vn{k}n\beta}
(\hat{\vn{M}}\times \delta \hat{\vn{M}})_{\beta}\Bigl\{ \\
&f_{\vn{k}n}{\rm Im}\sum_{m\neq n}
\frac{
\langle \psi_{\vn{k}n} |\vn{\mathcal{T}}|\psi_{\vn{k}m}\rangle
\langle \psi_{\vn{k}m} |[H_{\rm SOI},\sigma_{\beta}]|\psi_{\vn{k}n}\rangle
}{\mathcal{E}_{\vn{k}n}-\mathcal{E}_{\vn{k}m}}+\\
&+\frac{i}{2}
\delta(\mathcal{E}_{\rm F}-\mathcal{E}_{\vn{k}n})
\langle \psi_{\vn{k}n} |\vn{\mathcal{T}}|\psi_{\vn{k}n}\rangle
\langle \psi_{\vn{k}n} |[H_{\rm SOI},\sigma_{\beta}]|\psi_{\vn{k}n}\rangle\Bigr\}.
\end{aligned}\raisetag{4\baselineskip}
\end{gather}
Eq.~\eqref{eq_tmae_final} is well-suited for the calculation of
the magnetocrystalline anisotropy within Kohn-Sham density-functional-theory
codes. In contrast, the direct application of Eq.~\eqref{eq_mae1} in practice
would suffer from the following disadvantage:
Since the magnetocrystalline anisotropy energy is usually 
much smaller than the average exchange field
$\bar{\Omega}^{\rm xc}$,
one would need to calculate both $\bar{\Omega}^{\rm xc}$ as well 
as the torque-torque correlation function
$G^{\rm R}_{\mathcal{T}_{\alpha}\mathcal{T}_{\beta}}$ with very high precision 
if one wanted to use directly Eq.~\eqref{eq_mae1}
for the determination of the magnetocrystalline anisotropy.

In the absence of SOI we have $H_{\rm SOI}=0$ and 
Eq.~\eqref{eq_torque_soi_comm} simplifies to
$\mathcal{T}_{\beta}=i[H,\sigma_{\beta}]/2$. 
Since $|\psi_{\vn{k}n}\rangle$ is an eigenstate of $H$ it follows that
$\langle\psi_{\vn{k}n}|\mathcal{T}_{\beta}|\psi_{\vn{k}n}\rangle=0$
and therefore the last term in 
Eq.~\eqref{eq_retarded_torque_torque} vanishes. Thus, in the
absence of SOI Eq.~\eqref{eq_retarded_torque_torque} can be written as
\bege\label{eq_gtt_nosoi}
G^{\rm R}_{\mathcal{T}_{\alpha}\mathcal{T}_{\beta}}=
\frac{\hbar}{\mathcal{N}}\sum_{\vn{k}n}f_{\vn{k}n}{\rm Im}
\langle
\psi_{\vn{k}n}|\mathcal{T}_{\alpha}\sigma_{\beta}|\psi_{\vn{k}n}
\rangle.
\ee
Using $\sigma_{\alpha}\sigma_{\beta}=\delta_{\alpha\beta}+
i\epsilon_{\alpha\beta\gamma}\sigma_{\gamma}$ one can derive 
Eq.~\eqref{eq_gtt} 
from Eq.~\eqref{eq_gtt_nosoi}.

\end{document}